\documentclass[CRPHYS,Unicode,manuscript]{cedram}

\usepackage[margin=1in]{geometry}
\newcommand{\qr}{Q_{\textrm{\small{R}}}}
\newcommand{\qir}{Q_{\textrm{\small{IR}}}}
\newcommand{\qirmax}{Q_{\textrm{\small{IR,max}}}}
\newcommand{\qiro}{Q_{\textrm{\small{IR$_1$}}}}
\newcommand{\qirt}{Q_{\textrm{\small{IR$_2$}}}}
\newcommand{\qp}{Q_{\textrm{\small{P}}}}

\newcommand{\omir}{\Omega_{\textrm{\small{IR}}}}
\newcommand{\omr}{\Omega_{\textrm{\small{R}}}}
\newcommand{\aasqamu}{\AA$\sqrt{\textrm{amu}}$}
\newcommand{\qlx}{Q_{\textrm{\small{l$_x$}}}}
\newcommand{\qlz}{Q_{\textrm{\small{l$_z$}}}}
\newcommand{\qhx}{Q_{\textrm{\small{h$_x$}}}}
\newcommand{\omlx}{\Omega_{\textrm{\small{l$_x$}}}}
\newcommand{\omly}{\Omega_{\textrm{\small{l$_y$}}}}
\newcommand{\omlz}{\Omega_{\textrm{\small{l$_z$}}}}
\newcommand{\omhx}{\Omega_{\textrm{\small{h$_x$}}}}
\newcommand{\omhy}{\Omega_{\textrm{\small{h$_y$}}}}
\newcommand{\omiro}{\Omega_{\textrm{\small{IR$_1$}}}}
\newcommand{\omirt}{\Omega_{\textrm{\small{IR$_2$}}}}
\newcommand{\qwu}{Q_{2u}}
\newcommand{\qru}{Q_{3u}}

\newcommand{\bou}{B_{1u}}
\newcommand{\btu}{B_{2u}}
\newcommand{\bru}{B_{3u}}
\newcommand{\bog}{B_{1g}}

\NoLogo
\title{Light-control of materials via nonlinear phononics}

\author{\firstname{Alaska} \lastname{Subedi$^{1,2}$}}
\address{${}^{1}$CPHT, CNRS, Ecole Polytechnique, IP Paris, F-91128 Palaiseau, France}
\address{${}^{2}$Coll\`ege de France, 11 place Marcelin Berthelot, 75005 Paris, France}
\begin{abstract} 

Nonlinear phononics is the phenomenon in which a coherent dynamics in a
material along a set of phonons is launched after its infrared-active 
phonons are selectively excited using external light pulses.  The 
microscopic mechanism underlying this phenomenon is the nonlinear coupling
of the pumped infrared-active mode to other phonon modes present in a material.
Nonlinear phonon couplings can cause finite time-averaged atomic displacements 
with or without broken crystal symmetries depending on the order, magnitude 
and sign of the nonlinearities.  Such coherent lattice displacements along
phonon coordinates can be used to control the physical properties of materials
and even induce transient phases with lower symmetries.  Light-control of 
materials via nonlinear phononics has become a practical reality due to the 
availability of intense mid-infrared lasers that can drive large-amplitude 
oscillations of the infrared-active phonons of materials.  Mid-infrared pump 
induced insulator-metal transitions and spin and orbital order melting have been
observed in pump-probe experiments.  First principles based microscopic 
theory of nonlinear phononics has been developed, and it has been used to
better understand how the lattice evolves after a mid-infrared pump excitation
of infrared-active phonons.  This theory has been used to predict
light-induced switching of ferroelectric polarization as well as ferroelectricity in 
paraelectrics and ferromagnetism in antiferromagnets, which have been partially
confirmed in recent experiments.  This review summarizes the experimental and 
theoretical developments within this emerging field.

\end{abstract}

\begin{document}

% Use the \maketitle command after the abstract
\maketitle

%% Beginning of text
%% Abridged versions
%% 1. English one if the paper is in French
% \selectlanguage{english}
% \section*{Abridged English version}
% <your text here>
% \selectlanguage{french}
% to go back to main language.

%% 2. French one if the paper is in English
%  \selectlanguage{french}
%  \section*{Version française abrégée}
% Nous serons brefs.
\selectlanguage{english}
%to go back to main language.

\section{Introduction}

Light is a popular probe that is widely used to investigate the
structure and properties of materials. However, light has seldom been
used to coherently control materials, notwithstanding its use as a
source of heat.  New methods for modifying crystal structures
can lead to previously unexplored structures with unusual physical 
properties. Therefore, there is much interest in developing a technique
to control materials using light in the hope that this can be used
to stabilize hitherto unknown structures that have not been accessed
using pressure, heterostructuring or isovalent chemical doping.

On the practical side, light-control of materials requires intense
laser sources that excite materials and examine the excited state.
Such pump-probe laser setups were pioneered in chemistry laboratories,
and their development was instigated by the dream of controlling
chemical reactions by selectively exciting bond-specific vibrations
using ultrashort laser pulses \cite{zewa80}.  
In a thermodynamically activated chemical reaction, statistical laws 
imply that indiscriminately imparted energy to the reactants cause large
amplitude atomic vibrations along the weak bonds, which then get 
ruptured.  By using an intense laser pulse to induce large amplitude 
vibrations along a stronger bond-stretching mode, it has been shown 
that it is possible to modify the chemical reaction pathway and obtain 
a different set of products \cite{zare98}.  However, bond-selective 
light-control of chemical reactions has only been successful in relatively 
simple molecules with few atoms \cite{warr93}.  In complex molecules, 
it has been found that the vibrational energy in the pumped mode is quickly 
redistributed to other modes present in the molecule, and the pumped
mode dissipates and dephases before the amplitude of its vibration
becomes large \cite{bloe84}.

Vibrational energy redistribution occurs due to nonlinear couplings
between the vibrational modes of a molecule \cite{parm83,nesb96}.  
Two-dimensional (2D) pump-probe spectroscopy techniques have been 
developed that can simultaneously measure the oscillations of the 
pumped mode and other modes that are nonlinearly coupled to it \cite{wout02}.  
This has allowed the determination of the nature and magnitude of the nonlinear 
couplings, as well as the dissipation and dephasing times of the vibrations.
The nonlinear couplings between different vibrational modes appear as
off-diagonal peaks in 2D pump-probe spectroscopy measurements. In addition
to the lowest-order $Q_1Q_2$ coupling between two vibrational mode 
coordinates $Q_1$ and $Q_2$, higher-order $Q_1^{}Q_2^2$ nonlinearity 
has also been inferred from the measurements \cite{khal01,khal03}.  It has
also been noted that $Q_1^{}Q_2^2$ nonlinearity causes Frank-Condon-like 
displacement along the vibrational coordinates in the excited state 
\cite{huse03}.

Although nonlinear couplings between vibrational modes are an
impediment to bond-selective chemistry in large molecules, they do
lead to observation of coherent vibrations of nonlinearly coupled
modes in 2D pump-probe spectroscopy.  The discussion of a similar
effect in crystalline solids due to nonlinear coupling between phonons
starts with the proposal of ionic Raman scattering in the 1970s 
\cite{wall71,mart74}.  These studies showed that infrared-active (IR-active) phonons can 
play the role of an intermediate state in a Raman scattering process in the 
presence of a $\qr^{}\qir^2$ nonlinear coupling between a Raman-active 
phonon mode $\qr$ and an IR-active phonon mode $\qir$. 

Ionic Raman scattering has not yet been
observed in light scattering experiments.  However, F\"orst \textit{et
  al.}\ observed coherent oscillations at frequencies corresponding to
Raman-active phonon modes in their time-resolved reflectivity measurements
after a mid-IR pump in metallic
La$_{0.7}$Sr$_{0.3}$MnO$_3$ \cite{fors11}.  They proposed that these
oscillations occur because Raman-active modes
are coherently excited when an IR-active phonon mode is
externally pumped due to $\qr^{}\qir^2$ nonlinearities, and
this phenomenon has been called stimulated ionic Raman scattering.
In that study, oscillations of the pumped mode were not measured via
time-resolved spectroscopy experiments to show that
IR-active phonon excitations, and not electronic excitations, are 
responsible for the coherent oscillations of the Raman-active phonons. 
However, coherent oscillations at Raman-active phonon frequencies have
been observed in insulating ErFeO$_3$ \cite{nova16} and LaAlO$_3$ \cite{hort20} in
time-resolved spectroscopy experiments after a mid-IR pump, and
these experiments do show that the amplitude of the oscillations are
largest when the pump frequency is tuned to the frequency of the
IR-active phonon mode.  
Moreover, oscillations of the pumped
mode as well as the nonlinearly coupled low-frequency mode have been simultaneously
observed after a mid-IR pump in time-resolved second harmonic generation (SHG) experiment on
LiNbO$_3$ \cite{mank17}, which conclusively demonstrates the phenonmenon of
stimulated ionic Raman scattering.

F\"orst \textit{et al.}\ have also noted that a Raman-active phonon mode
experiences a force proportional to $g \qir^2$ in the presence of
a nonlinear coupling term $g \qr^{}\qir^2$ with coupling constant $g$ \cite{fors11}.  
Since the force is
proportional to the square of the IR-active phonon coordinate, total
force exerted on the Raman-active mode has a nonzero time-averaged value while
the IR-active mode is oscillating. This causes the lattice to get displaced along
the Raman-active phonon coordinate when the IR-active mode is pumped, and this
phenonmena has been termed nonlinear phononics.  Time-resolved x-ray
diffraction experiments on La$_{0.7}$Sr$_{0.3}$MnO$_3$ and
YBa$_2$Cu$_3$O$_{6.5}$ after an intense mid-IR excitation have
found intensity modulation of Bragg peaks of less than 0.5\%, and it has been
argued that these modulations are due to lattice displacement along Raman-active
phonon coordinates \cite{fors13,mank14}.  In
YBa$_2$Cu$_3$O$_{6.5}$, this corresponds to bond length changes of
less than 1 pm. Furthermore, these materials are metallic, and
oscillations of the pumped mode have not been measured in these
experiments to rule out electronic excitation as a cause of the
structural changes.
More convincing evidence of light-induced displacement due to nonlinear
phononics has been proposed \cite{sube15} and then observed in ferroelectric
LiNbO$_3$, which has a high-frequency IR-active phonon with a large oscillator
strength \cite{mank17}. In this material, a strong reduction and
sign reversal of the electric dipole moment and a simultaneous oscillation at
the frequency of the pumped IR-active mode has been observed in pump-probe 
SHG experiments. This demonstrates that a coherent displacement of the
lattice along a phonon coordinate is feasible at least in insulators
that have IR-active phonons with a large oscillator strength.

Coherent lattice displacements due to nonlinear phononics after
mid-IR excitations have been attributed to be the cause of insulator-metal
transitions in Pr$_{1-x}$Ca$_{x}$MnO$_3$ ($x = 0.3, 0.5$) \cite{rini07,espo17} and NdNiO$_3$ \cite{cavi12}, 
orbital order melting in La$_{0.5}$Sr$_{1.5}$MnO$_3$ \cite{tobe08,fors11b}, and 
magnetic order melting in NdNiO$_3$ \cite{fors15a}.  Although phase transitions 
in these materials occur after a mid-IR pump, neither coherent lattice displacements
nor excitations of the pumped IR-active phonon was demonstrated in any of
these experiments.  Therefore, it is not yet known if lattice displacements
due to nonlinear phononics can be large enough to modify the physical properties
of materials. Most of the currently reported phase transitions due to mid-IR
excitations involve melting of order.  Since light pulses always impart heat, 
it may be infeasible to unambiguously show that melting of order 
is due to nonlinear phononics because it might not be possible to disentangle
the effects of heating and light-induced phonon excitations.
Light-control of materials properties via nonlinear phononics can only
be conclusive when it involves breaking of symmetries that are present in the
equilibrium phase.  Light-induced breaking of inversion symmetry in oxide 
paraelectrics has been theoretically predicted \cite{sube17}, and Nova 
\textit{et al.}\ have observed metastable ferroelectricity in SrTiO$_3$ after
a mid-IR pump \cite{nova19}.  However, the same effect in SrTiO$_3$ has also been achieved
using terahertz pump \cite{li19}.  So it is not yet clear if the observed mid-IR 
pump induced ferroelectricity is caused by lattice displacements due to nonlinear
phononics.

Mid-IR pump terahertz probe experiments have been performed on several 
superconductors \cite{faus11,kais14,hu14,mitr16}.  The 
reflected electric field transients of probe pulses are enhanced by a few 
percent in these experiments, and the nonequilibrium state with enhanced 
reflectivity relaxes to the normal state within 1--2 ps.  Although the 
low-frequency optical conductivity of a short-lived state is not a well 
defined quantity, the Fourier transform of the reflected electric field 
transients have been analyzed in terms of frequency-domain optical conductivity.  
The increased time-dependent reflectivity after a mid-IR pump appears as a 
low-frequency peak in the imaginary part of the reconstructed optical 
conductivity $\sigma_2(\omega)$, and this has been interpreted as a signature of 
light-induced superconductivity. In a true superconducting state, the real
part of the optical conductivity $\sigma_1(\omega)$ should concomitantly decrease at low 
frequencies.  However, the reconstructed $\sigma_1(\omega)$ shows an increase
at low frequencies.
Moreover, there have been no two-dimensional spectroscopy experiments
to show that an IR-active phonon mode is actually being pumped while the 
metastable state with increased reflectivity gets transiently realized, making the connection
of nonlinear phononics phenomena in these experiments rather tenuous.  In fact,
the reconstructed $\sigma_2(\omega)$ shows a superconductivity-like behaviour 
after a mid-IR pump in samples where even the IR-active phonon is not observed
in optical spectroscopy experiments \cite{mitr16}, suggesting that the observed light-induced effect
is due to electronic excitations.
Since light-induced superconductivity is not apparent in the raw data and only 
manifests after data analysis, this subject will not be discussed in this review.
Readers interested in this subject are pointed to reviews that discuss the 
mid-IR pump experiments in superconductors in detail \cite{mank16,nico16,cava18}.

% development of the theory of nonlinear phononics

Light-induced structural dynamics in crystals has been theoretically
studied using molecular dynamics \cite{qi09} and time-dependent density functional
theory calculations \cite{shin10}.  These methods have the advantage of taking into
account the light-induced changes in all the dynamical degrees of freedom 
present in the material.  But it is
cumbersome to extract the relevant nonlinear couplings between the
pumped IR-active mode and coupled Raman-active modes from these methods,
which hinders the understanding of the microscopic processes that cause
the light-induced phenomena.
Subedi \textit{et al.}\ started the use of a theoretical framework to
study nonlinear phononics that is based on symmetry principles to 
identify the symmetry-allowed nonlinear couplings, first principles
calculations of the coefficients of these nonlinear terms, and
solution of the equations of the motions for the coupled phonon
coordinates \cite{sube14}.
This framework was used to explain the light-induced phenonmena
observed in the pioneering mid-IR pump experiments in the manganites
 \cite{rini07,fors11}.  Calculations based on this framework was used
to reconstruct the mid-IR pump-induced changes in the structure of 
YBa$_2$Cu$_3$O$_{6.5}$ \cite{mank14} and explain the observation of 
symmetry-breaking Raman-active modes in orthoferrites \cite{jura17a}. 
A mechanism for ultrafast switching of ferroelectricity was proposed
using this method \cite{sube15}, and a recent observation of momentary
reversal of ferroeletricity in LiNbO$_3$ has partially confirmed this 
prediction \cite{mank17}.  Light-induce 
ferroelectricity \cite{sube17} and ferromagnetism \cite{rada18} have 
also been proposed, the latter prediction based solely on symmetry 
arguments.  Recent mid-IR pump experiments have observed long-lived
metastable ferroelectricity in SrTiO$_3$ \cite{nova19} and ferromagnetism
in DyFeO$_3$ \cite{afan19} and  CoF$_2$ \cite{disa20}, but the precise mechanisms for these phenomena
need to be clarified with further experimental studies.  Other theoretical 
predictions based on the nonlinear phonon couplings that await experimental
confirmations include indirect-to-direct band gap switching \cite{gu16},
phono-magnetic analogs of opto-magnetic effects \cite{jura19a}, cavity
control of nonlinear phonon interactions \cite{jura19b}, and excitation of an
optically silent mode in InMnO$_3$ \cite{jura20}.

Recent successes in light-control of materials using mid-IR pulses show
that nonlinear phononics has an important role to play at the frontier of
materials physics.  This is further underscored by the construction of several free electron laser 
sources that have recently come online or are in the process of being built.
There have been several reviews of this field that have focused on the
experimental aspects of the field \cite{mank16,nico16,cava18,fors15,buzz19,sale19}.  
This review attempts to summarize the experimental and theoretical developments
in the field of nonlinear phononics, emphasizing how theoretical calculations
have helped the experimentalists drive this field forward.

\section{Theoretical approach}
\label{}

In nonlinear phononics experiments, experimentalists are equipped with a light
source that can strongly excite some set of IR-active phonons of a material, and
they want to understand how the structure and physical properties of
the material changes after the phonon excitation. 
In another situation, experimentalists know a phonon mode associated with an
order parameter, and they want to identify the IR-active phonons that should be
pumped to alter the ordered state by coherently displacing the lattice along
the phonon coordinate of the order parameter. 
Subedi \textit{et al.}\ initiated the use of a microscopic theory
to quantify the nonlinear couplings between the IR- and Raman-active phonons and 
predict the light-induced structural dynamics from first principles 
\cite{sube14}.
This framework relies on i) symmetry principles to determine
which phonon modes can couple to the pumped IR-active phonon, ii) first
principles calculation of the energy surface as a function of the coupled
phonon coordinates to extract their nonlinear couplings, and iii)
solution of the coupled equations of motions for the phonon
coordinates.

The first step in a nonlinear phononics experiment is the identification of
phonons of a material.  Since light couples to phonons near the Brillouin
zone center, the phonons relevant to a nonlinear phononics experiment are
measured using optical reflectivity and Raman scattering experiments. In 
addition to measuring the frequencies, these experimental methods also
yield information about the symmetry of the phonons.  In the theory side, density
functional theory (DFT) based methods can 
calculate forces on atoms, and these can be used to construct dynamical 
matrices.  Diagonalization of the dynamical matrices produces phonon frequencies
and their irreducible representation (irrep) can be determined by studying how 
the corresponding eigenvectors
transform under the symmetry operations of the point group of the material. 
In this way, the phonon frequencies and their symmetries can be reconciled between
experiment and theory.

The ability to pump a particular set of phonons in a material depends on the
frequency and power of the available laser source.  The pump-induced response of
a material is usually investigated by analyzing the reflected, transmitted or scattered 
probe pulses.  The phonons that are nonlinearly coupled to the pumped IR-active 
mode are observed as oscillations in the amplitude of the detected probe pulse.  
Although the response along the coupled phonon modes can easily be detected after a
pump, extraction of nonlinear couplings from experimental data has so far 
proven to be difficult.  Additionally, coherent change in lattice can also be
inferred from time-resolved diffraction experiments, but a complete determination
of the structural changes has been impractical because this requires measuring 
changes in numerous diffraction peaks as a function of time after a pump.  The
usefulness of the DFT-based microscopic theory of nonlinear phononics lies in
the ability to calculate nonlinear couplings from first principles, which makes
it possible to efficiently calculate and predict pump-induced structural changes.

To extract the nonlinear phonon couplings between two modes, the calculated
phonon eigenvectors are used to generate a large number of structures as a
function of the two phonon coordinates.  DFT calculations are run again to
calculate total energies of these structures.  The total energies are then collected
and fit with a polynomial to extract the coefficients corresponding to the
nonlinear terms in the expansion of the total energy as a function of the phonon
coordinates.  Although a generic polynomial can be used to fit the calculated 
total energy surface, polynomials with only symmetry-allowed terms are used in
the fitting to reduce the sources of numerical noise.  The symmetry allowed
terms are determined by noting that the total energy is a scalar and has the
trivial irrep.  For the cubic order coupling $\qr^{}\qir^2$, it implies that an 
IR-active mode can only couple to a Raman mode with the irrep $A_g$ in
centrosymmetric crystals because the square of an irrep is the trivial irrep
$A_g$.  However, if two different IR modes are pumped simultaneously, then their
coupling to the $A_g$ mode is forbidden with the $\qr\qiro\qirt$ term because
the product of two different odd irreps cannot equal the trivial irrep.  For
example, if phonons with the irreps $B_{2u}$ and $B_{3u}$ of a material with 
the point group $mmm$ are pumped simultaneously, only Raman-active modes with the 
irrep $B_{1g}$ can nonlinearly couple to them because $B_{1g} \subseteq B_{2u} \otimes
B_{3u}$.  For the quartic term $\qr^2\qir^2$, any Raman mode can couple to the
pumped mode because the square of any irrep is the trivial irrep.

After the nonlinear coefficients present in the polynomial fit of the 
calculated energy surface as a function of phonon coordinates are extracted,
the phonons are treated as classical oscillators to study their light-induced
dynamics.  The coupled equations of motion for phonon coordinates are
numerically solved in the presence of a forcing term of the form $F(t) =
Z_{\alpha}^*F_0 \sin(\Omega t) e^{-t^2/\sigma^2}$. Here, $Z_{\alpha}^*$ is the 
component of the mode effective charge along the direction $\alpha$, and it is
related to the Born effective charge tensor $Z_{\kappa, \alpha \beta}^*$ of 
atom $\kappa$ with mass $m_\kappa$ and the eigendisplacement vector
$w_{\kappa, \beta}$ by the expression $Z_{\alpha}^* = \Sigma_{\kappa,\beta} 
Z_{\alpha \beta}^* w_{\kappa, \beta} / \sqrt{m_{\kappa}}$ \cite{sube17}.  
The mode effective charge can be calculated from first principles and is related
to the LO-TO splitting that can be experimentally measured.  Hence, except for 
the damping of the phonon modes that are roughly 10--20\% of the respective
phonon frequencies, all the quantities necessary for calculating the
light-induced dynamics can be determined from first principles.

\section{Coherent lattice displacement due to cubic-order coupling}
\label{}

\subsection{Coherent displacement in Pr$_{0.7}$Ca$_{0.3}$MnO$_3$}
% Cubic $a_{12}\qr\qir^2$ nonlinear coupling was discussed in the 1970s
% as a potential mechanism for ionic Raman effect. F\"orst \textit{et
%   al.}\ pointed out in 2011 that this coupling can be used to
% coherently displace crystals along the $\qr$ phonon coordinate while
% the $\qir$ mode is being pumped.

An insulator-metal transition in Pr$_{0.7}$Ca$_{0.3}$MnO$_3$ after an 
excitation with a mid-IR laser was reported by Rini \textit{et al.}\ in 2007 
\cite{rini07}.  This was the first report of a pump-probe experiment in a 
transition metal oxide using a mid-IR pump that could excite the high-frequency 
IR-active phonons 
of the material causing changes in the metal-oxygen bond distances.  The
observed insulator-metal transition was understood in terms of the modification
of the electronic bandwidth associated with the changes in these bond distances.
In 2012, F\"orst \textit{et al.}\ proposed that nonlinear phonon coupling 
of the type $\qr^{}\qir^2$ can cause a coherent lattice displacement along the 
Raman-active phonon coordinate $\qr$ when the IR-active phonon coordinate $\qir$
is externally pumped \cite{fors11}.  Subedi 
\textit{et al.}\ theoretically investigated whether this nonlinear phononics 
phenonmenon can explain the observed insulator-metal transition in 
Pr$_{0.7}$Ca$_{0.3}$MnO$_3$ and found that externally exciting an IR-active 
phonon mode of this material can cause a lattice displacement along a low-frequency
Raman-active phonon coordinate \cite{sube14}. 

\begin{figure}
  \begin{center}
    \includegraphics[width=0.5\columnwidth]{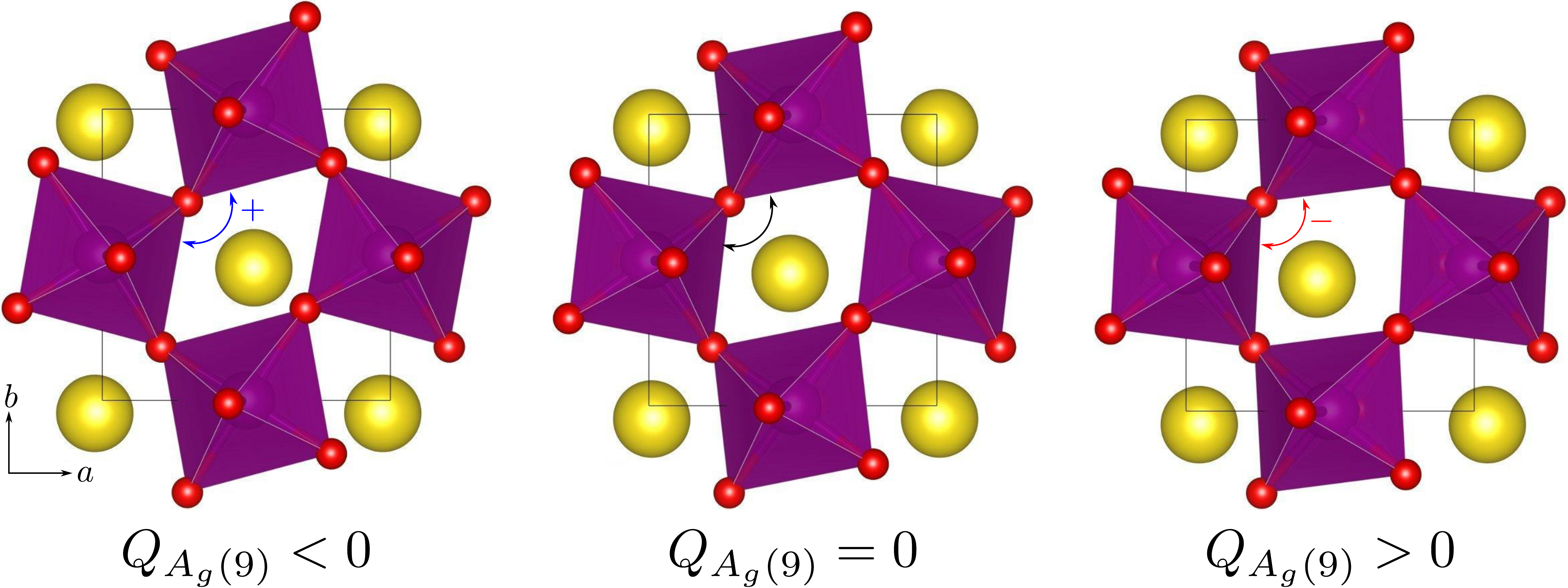}
    \includegraphics[width=0.5\columnwidth]{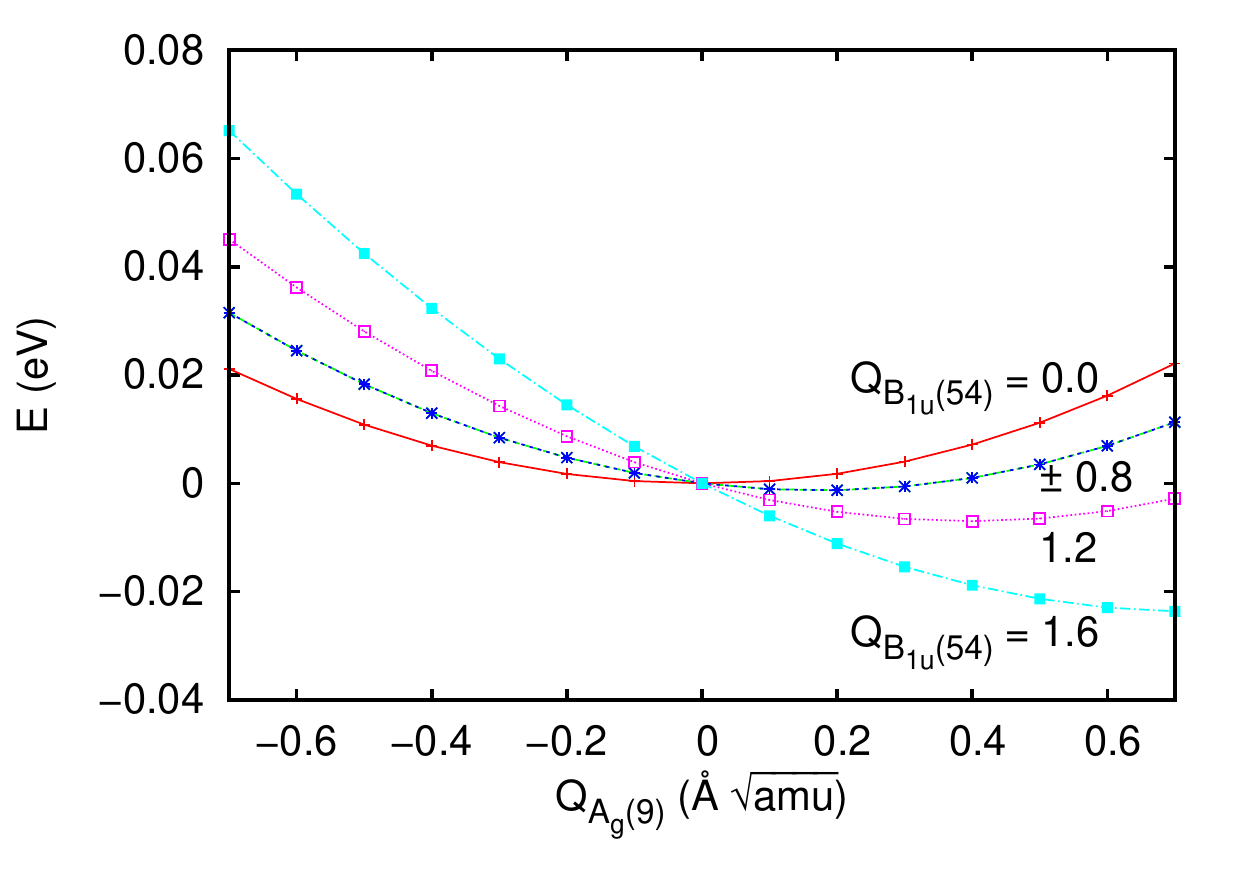}
    \end{center}
  \caption{Top: Sketch of the atomic displacements corresponding to the 
    $A_g(9)$ Raman-active phonon mode of PrMnO$_3$ A positive value of the
    $A_g(9)$ coordinate brings the angle between octahedra closer to
    90$^{\circ}$. Bottom: Total energy as a function of the $A_g(9)$ 
    coordinate for several values of the $B_{1u}$ coordinate. For visual
    clarity $E(\qr,\qir) - E(0,\qir)$ is plotted so that all curves coincide
    at $\qr = 0$.  Reproduced from \cite{sube14}.}
  \label{fig:pmo-ene}
\end{figure}

Optical spectroscopy shows that there is an IR-active phonon mode with a
large oscillator strength in this material at 573 cm$^{-1}$ \cite{okim99}.
In their experiment, Rini \textit{et al.}\ used a mid-IR laser with a
frequency near 573 cm$^{-1}$ to pump this highest-frequency IR-active
phonon of the material.  For computational efficiency, Subedi
\textit{et al.}\ studied this phenomenon on the parent compound
PrMnO$_3$.  They started by calculating the zone center phonon
frequencies and eigenvectors of this material.  The calculated frequencies of the three
highest IR-active phonon modes are 633, 640, and 661 cm$^{-1}$ with
irreps $\bou$, $\btu$, and $\bru$, respectively.  Since the product
of an irrep with itself is the trivial irrep $A_g$, any of the seven
$A_g$ phonon modes present in the material can couple with an
IR-active mode with a cubic-order $\qr^{}\qir^2$ nonlinear
coupling.  The irreps of the pumped IR-active mode and the $A_g$ mode
that couples to the pumped mode was not reported in the experimental study
\cite{rini07}.  So
total energies $E(\qr,\qir)$ as a function of the $\qr$ and $\qir$
coordinates were calculated for each pair of $A_g$ and high-frequency
IR-active phonon modes.  The $A_g(9)$ and $\bou$ modes showed a large nonlinear
coupling, and the calculated energy surface of this
pair is shown in Fig.~\ref{fig:pmo-ene}(bottom).  The high-frequency $\bou$ mode
involves changes in the bond length between the apical O and Mn of the
MnO$_6$ octahedra.  The $A_g(9)$ mode has a relatively low calculated
frequency of 155 cm$^{-1}$ and involves rotation of the MnO$_6$
octahedra about the $c$ axis as shown in Fig.~\ref{fig:pmo-ene}(top). The
in-plane angle between the corner-shared octahedra become closer to
90$^{\circ}$ for positive values of the $A_g(9)$ coordinate, while
the distortion increases for negative values.

\begin{figure}
  \begin{center}
    \includegraphics[width=0.9\columnwidth]{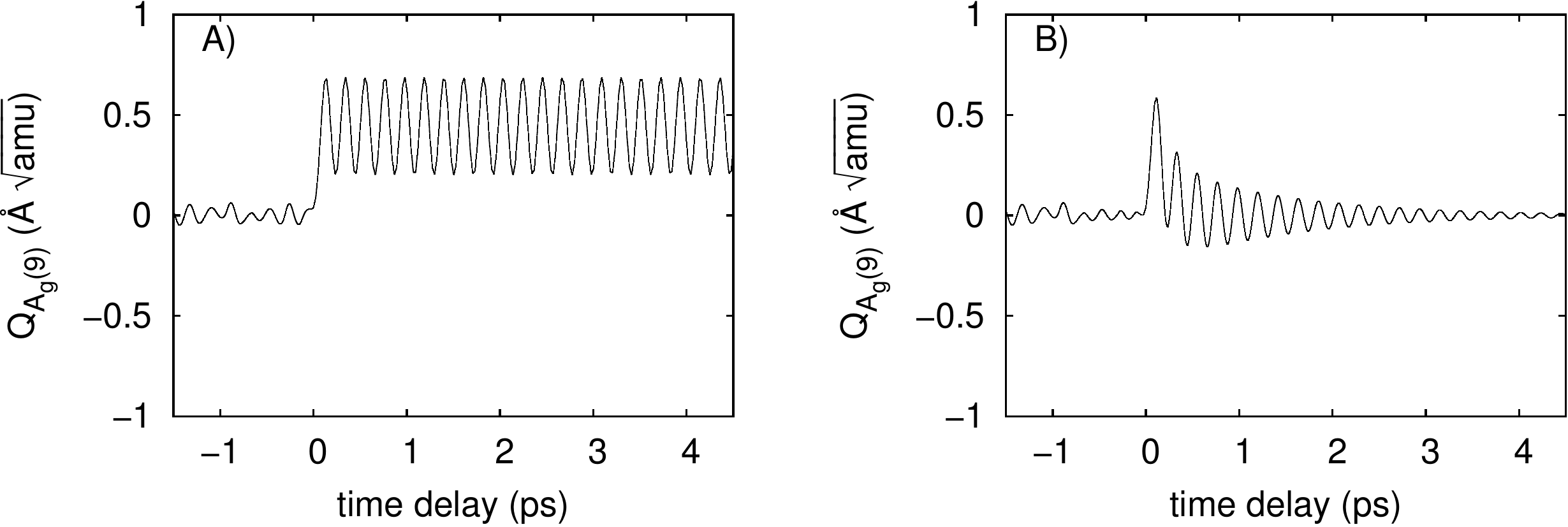}
    \end{center}
  \caption{Dynamics of the $A_g(9)$ Raman mode. Left panel: dynamics without damping.
    Right panel: dynamics with damping values of 5\% for both $A_g(9)$ and $B_{1u}$ 
    modes. Reproduced from \cite{sube14}.}
  \label{fig:pmo-dyn}
\end{figure}

The calculated energy surface between $A_g(9)$ and $\bou$ modes fits
the following polynomial
\begin{eqnarray}
E(\qr,\qir) & = & \frac{1}{2} \omr^2 \qr^2 + \frac{1}{2} \omir^2\qir^2 
+ \frac{1}{3} a_3 \qr^3 + \frac{1}{4} b_4 \qir^4
\nonumber \\ 
& &-\frac{1}{2} g \qr^{}\qir^2,
\label{eq:energ_pmo}
\end{eqnarray}
with $\qr$ and $\qir$ now denoting the coordinates of the $A_g(9)$ and
$\bou$ modes, respectively.  Since the pump-induced dynamics causes
large displacements along the phonon coordinates, they can be regarded
as classical oscillators.  The effect of an external pump on the
IR-active mode can be treated by the presence of a driving term
$F(t)=F\sin(\Omega t) e^{-t^2/2\sigma^2}$, where $F$, $\sigma$, and
$\Omega$ are the amplitude, time-width and frequency of the pump pulse,
respectively.  The zero of the time delay $t$ is typically chosen such 
that $t = 0$ when the pump and probe pulses are superposed.  The coupled 
equations of motions for the two phonon coordinates in the absence of 
damping terms then read
\begin{eqnarray}
\ddot{Q}_{\textrm{\small{IR}}}+\omir^2\qir^{}&=& g \qr\qir -  b_4 \qir^3 + F(t) \nonumber \\
\ddot{Q}_{\textrm{\small{R}}}+\omr^2\qr^{}&=& \frac{1}{2} g \qir^2 - a_3 \qr^2\,.
\end{eqnarray}
For a finite pump amplitude $F$, the oscillation of the $\qir$
coordinate has a time dependence $\qir(t) \propto F \omir \sigma^{3}
\cos \omir t$ \cite{sube14}.  Due to the cubic-order coupling between the two
modes, the Raman coordinate experiences a forcing field $g \qir^2/2
\propto g F^2\omir^2\sigma^6 (1-\cos 2\omir t)$, which has a finite
time-averaged value.  Therefore, unlike the pumped IR-active mode, the
Raman mode vibrates about a displaced position.  Numerical solution of
these coupled equations of motion also confirms this picture.
Fig.~\ref{fig:pmo-dyn}(a) and \ref{fig:pmo-dyn}(b) show the dynamics of the $\qr$ coordinate with
and without damping terms.  In both cases, the $\qr$ coordinate
oscillates about a displaced position while the $\qir$ mode is also
oscillating with a finite amplitude.

\begin{figure}[bh]
  \begin{center}
    \includegraphics[width=0.8\columnwidth]{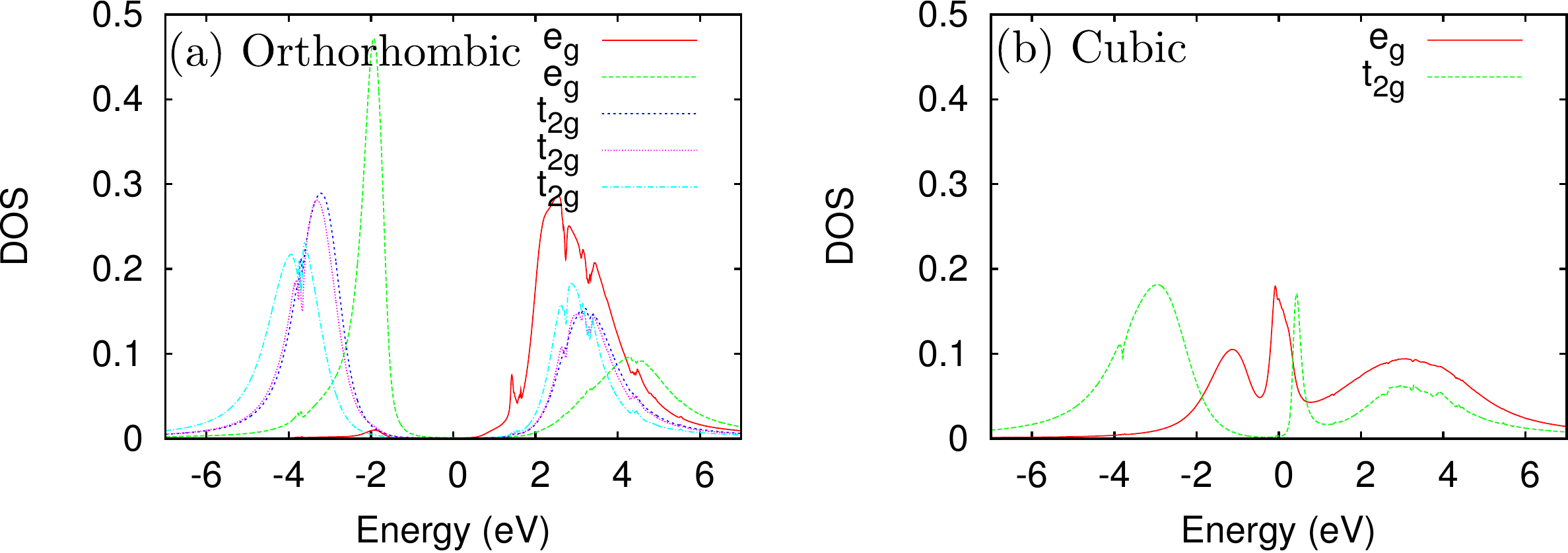}
    \end{center}
  \caption{DFT+DMFT orbitally resolved density of states of Mn-$d$ states in
  PrMnO$_3$ for the equilibrium orthorhombic (insulating) and cubic (metallic)
  crystal structures.  Reproduced from \cite{sube14}.}
  \label{fig:pmo-maxent}
\end{figure}

Since the lattice displaces along the positive value of the $A_g(9)$ coordinate while
 the $\bou$ is pumped,
the rotation of the MnO$_6$ octahedra in the $ab$
plane gets reduced. This should bring the system closer to the metallic phase
because reduced octahedral rotation enhances hopping of the Mn $d$ electrons
via oxygen sites.
Subedi \textit{et al.}\ performed combined density functional theory
and dynamical mean field theory (DMFT)
electronic structure
calculations on the cubic and equilibrium structures of PrMnO$_3$ to understand
how the electronic structure changes as the octahedral rotation is suppressed.
The calculated partial density of states of the Mn $t_{2g}$ and $e_g$ orbitals 
for the two structures are shown in Fig.~\ref{fig:pmo-maxent}.  They show that 
the distorted equilibrium structure is insulating, while the
cubic structure is metallic.  This suggests that light-induced suppression of the
octahedral rotation due to nonlinear phononics might cause
insulator-metal transition in this material. 

Has it been conclusively shown that the observed mid-IR pump-induced
insulator-metal transition in Pr$_{0.7}$Ca$_{0.3}$MnO$_3$ is due to
displacement of the lattice along the Raman coordinate?  
It is worthwhile to point out that any magnetic, optical, and electrical
perturbations easily cause insulator-metal transition in 
Pr$_{0.7}$Ca$_{0.3}$MnO$_3$ \cite{tomi96,miya97,asam97}.
There have been no experiments to measure the
oscillations of the pumped IR- and Raman-active modes using 2D
spectroscopy to show that these phonons are excited in the experiment, 
nor have the light-induced changes in the structure been
studied using time-resolved x-ray diffraction spectroscopy.  Moreover,
since the light-induced transition is from an insulating to a metallic
phase, heating effects cannot be ruled out as the cause of the
transition unless the timescale for thermal redistribution of the
pumped vibrational energy is disentangled from the timescale of any
displacement along the Raman coordinate.  As a result, it may be practically
impossible to conclusively prove that pump-induced insulator-metal transition
in this material is due to a displacement along the Raman coordinate.

\subsection{Coherent displacement in ortho-II YBa$_2$Cu$_3$O$_{6.5}$}

A mid-IR pump-induced increase in reflectivity has been observed in
several cuprates, and this fact from raw data has been interpreted as
a signature of light-induced transient superconductivity 
\cite{faus11,kais14,hu14}. Mankowsky \textit{et al.}\ have performed 
a combined time-resolved x-ray diffraction and
first-principles lattice dynamics study to find out if the
light-induced effect observed in YBa$_2$Cu$_3$O$_{6.5}$ is due to
structural changes caused by nonlinear phononics \cite{mank14}.  Optical
spectroscopy experiments show that this material has an IR-active
phonon with a frequency of 640 cm$^{-1}$ \cite{home95}.  This mode has the irrep
$\bou$.  Mankowsky \textit{et al.}\ pumped this mode with an intense
mid-IR laser pulse and measured changes in the diffraction intensity of four Bragg
peaks as a function of time using time-resolved x-ray diffraction experiment.
The intensities either increased or decreased promptly after a mid-IR
pump.  Since the intensity of a Bragg peak is proportional to the square of the 
structure factor,
which is a function of atomic positions, this implies that the crystal
structure of the material coherently changes after the pump.  However,
ortho-II YBa$_2$Cu$_3$O$_{6.5}$ has 25 atoms, and they were
not able to fully resolve the light-induced changes in the crystal
structure by measuring the changes in intensities of only four Bragg peaks.

\begin{figure}
  \begin{center}
    \includegraphics[width=0.4\columnwidth]{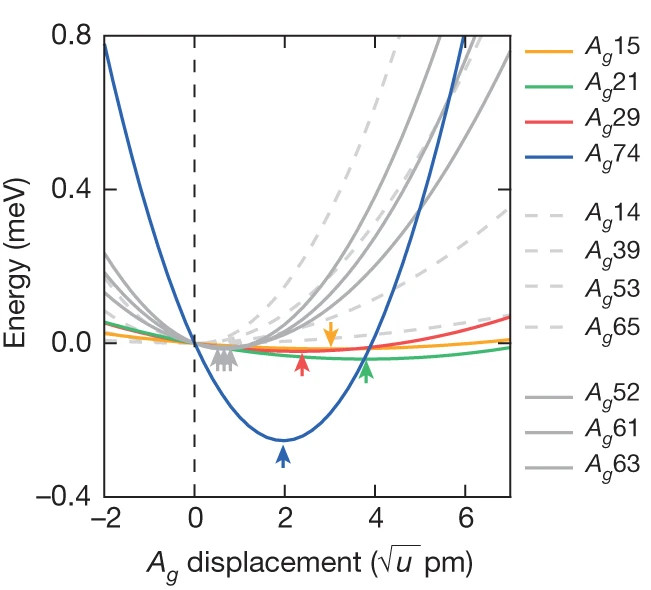}
    \end{center}
  \caption{Calculated total energy of all the $A_g$ modes for a frozen $B_{1u}$ 
    displacement of 0.14 \AA $\sqrt{u}$ ($u$, atomic mass unit), corresponding
    to a change in the apical O--Cu distance of 2.2 pm.  Arrows indicate the potential 
    minima. Reproduced from \cite{mank14}.}
  \label{fig:ybco-pot}
\end{figure}

\begin{figure}
  \begin{center}
    \includegraphics[width=0.7\columnwidth]{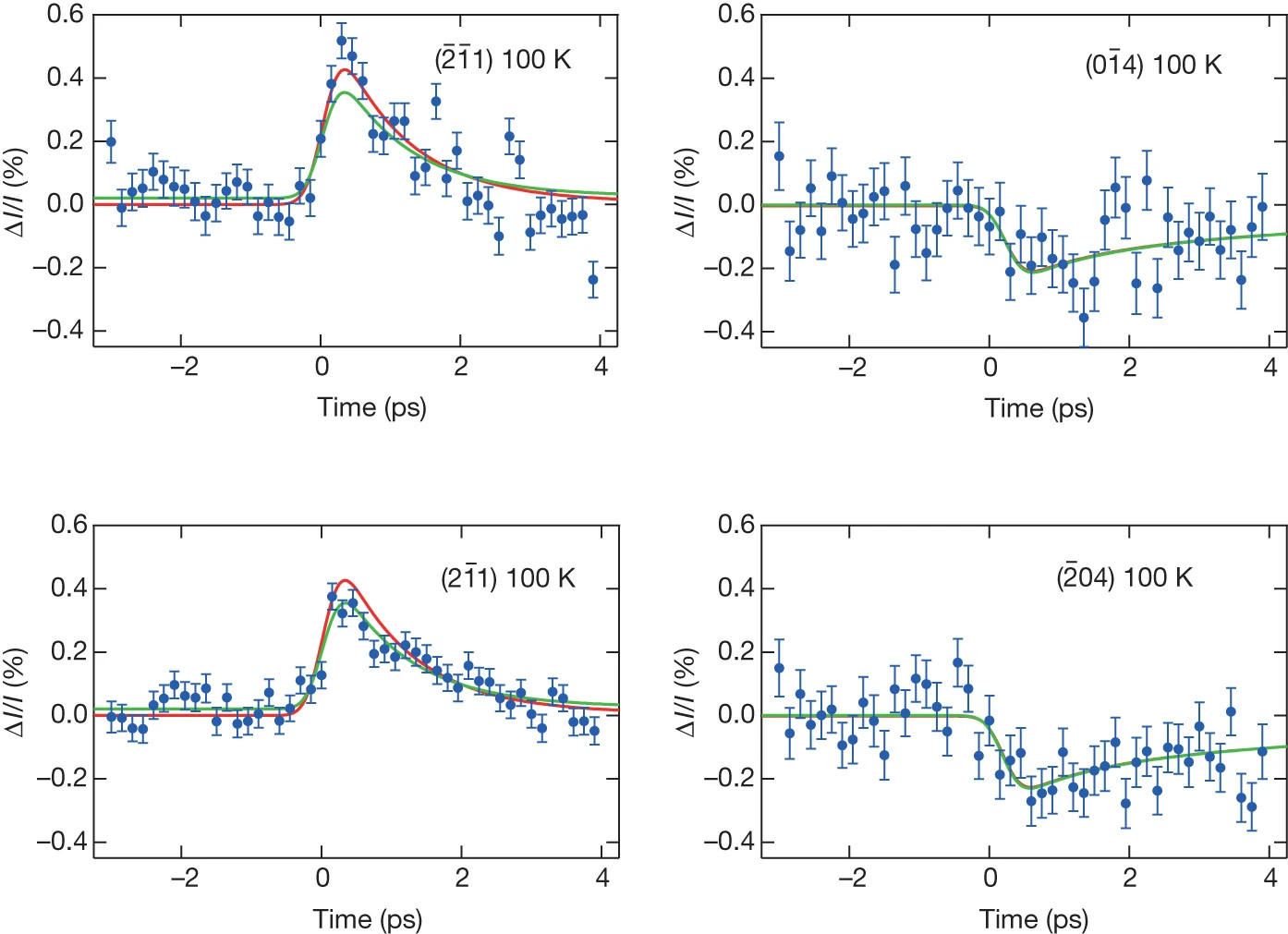}
    \end{center}
  \caption{Time-dependent diffracted peak intensity ($I$) for four Bragg reflections.
     The solid curves are fit to the experimental data which were done by adjusting
     the $B_{1u}$ amplitude and relaxation times. The relative amplitudes and signs 
     of the curves are determined from the calculated structure using only the four 
     most strongly coupled modes (green) or all $A_g$ modes (red).  Reproduced from
     \cite{mank14}.}
  \label{fig:ybco-braggs}
\end{figure}

There are 11 $A_g$ modes in this material, and all of them can couple
to the pumped $\bou$ mode with a cubic-order $\qr\qir^2$
coupling.  Mankowsky \textit{et al.}\ calculated the total
energy surface for each pair of $\bou$ and $A_g$ modes.  The total
energies of the seven $A_g$ modes for a $\bou$ displacement of 0.14
\aasqamu\ are shown in Fig.~\ref{fig:ybco-pot}.  The calculated energy curves show
that four $A_g$ modes
show significant coupling to the pumped $\bou$ mode, and these modes involve
out-of-plane motion of the apical O and Cu ions.  The rest of the
$A_g$ modes that are weakly coupled involve in-plane motions of the O
ions in the CuO$_2$ plane.  The presence of these nonlinearities was also 
independently confirmed by the calculations of Ref.~\cite{fech16}.
Changes in the intensities of the four Bragg
peaks measured in the time-resolved x-ray diffraction experiment 
were calculated considering the displacement of the lattice
along these $A_g$ coordinates. 
The measured and calculated changes in the intensities of the Bragg peaks as a
function of time are shown in Fig.~\ref{fig:ybco-braggs}.
With only the $\bou$ pump amplitude and
the decay time as the fitting parameters, the changes in the crystal structure
due to displacement along the $A_g$ coordinates could independently
reproduce the pump-induced changes in intensities of the four measured Bragg peaks.
In the transient structure corresponding to the $\bou$ amplitude of 0.3 
\AA$\sqrt{\textrm{amu}}$ estimated for the pump intensity utilized in the the experiment,
the apical O-Cu distance decreases and the O-Cu-O buckling increases.  There is also an
increase of the intra-bilayer distance and a decrease of the
inter-bilayer distance.  The changes in the distances are around 1 pm,
and DFT calculations show that these cause practically no modification
of the electronic structure for the estimated pump-induced amplitude
of the $\bou$ mode in the experiment \cite{mank14}. This suggests that structural changes 
due to nonlinear phononics do not cause the observed light-induced enhancement 
of reflectivity in this material.

% Since the light-induced changes
% in the structure are so small, other experimental techniques such as 2D spectroscopy should be
% performed to confirm that displacements along the $A_g$ coordinates are
% caused by pumping the $\bou$ mode.  

\subsection{Excitation of Raman modes with nontrivial irreps in ErFeO$_3$}

ErFeO$_3$ is an insulator with a band gap of 2.1 eV, and it shows
resonances at 540 and 567 cm$^{-1}$ in the optical conductivity spectra
corresponding to phonons with irreps $\bru$ and $\btu$, respectively \cite{subb70}.
When either the $\bru$ or $\btu$ mode of this material was externally pumped with
light polarized along $a$ or $b$ axes, respectively, Nova
\textit{et al.}\ observed oscillations at the frequency of 112 cm$^{-1}$ 
corresponding to an $A_g$ Raman-active phonon mode \cite{nova16}. 
This reflects the
cubic-order $\qr^{}\qir^2$ coupling between the pumped IR-active
and $A_g$ modes.  In addition, they measured two $\bog$ phonons with
frequencies of 112 and 162 cm$^{-1}$ when the $\bru$ and $\btu$ modes
were simultaneously pumped.  The excitation of the $\bog$ phonons is
due to a cubic-order $Q_{B_{1g}} Q_{B_{2u}} Q_{B_{3u}}$
coupling that is allowed by symmetry because $B_{1g} \subseteq B_{2u}
\times B_{3u}$.  Juraschek \textit{et al.}\ studied the dynamics of
these phonons in ErFeO$_3$ using theoretical framework described above and found large
symmetry-allowed cubic-order couplings between the Raman- and IR-active phonons
 that explains the observed pump-induced Raman oscillations \cite{jura17a}.

The observation of stimulated oscillations of Raman phonons due to
nonlinear phonon couplings in ErFeO$_3$ is interesting because the
band gap of this material is large enough that the role of electronic
excitations in causing the light-induced dynamics can reasonably be
ruled out.  The observation of Raman oscillations at only two
frequencies also raises an interesting question.  Why are not the oscillations
of other Raman $A_g$ and $\bog$ modes observed?  It can be
conjectured that the respective nonlinear couplings are small or that
the energy from the pumped IR modes flows mostly to low frequency
Raman modes. It would be illuminating to perform experimental and theoretical
studies that can clarify this issue. It would also be interesting to perform
time-resolved x-ray diffraction experiment to find 
out whether the lattice displaces along the Raman-active phonon coordinates after a
mid-IR pump in this material.

\subsection{Transient switching of ferroelectricity}

\begin{figure}
  \begin{center}
    \includegraphics[width=0.5\columnwidth]{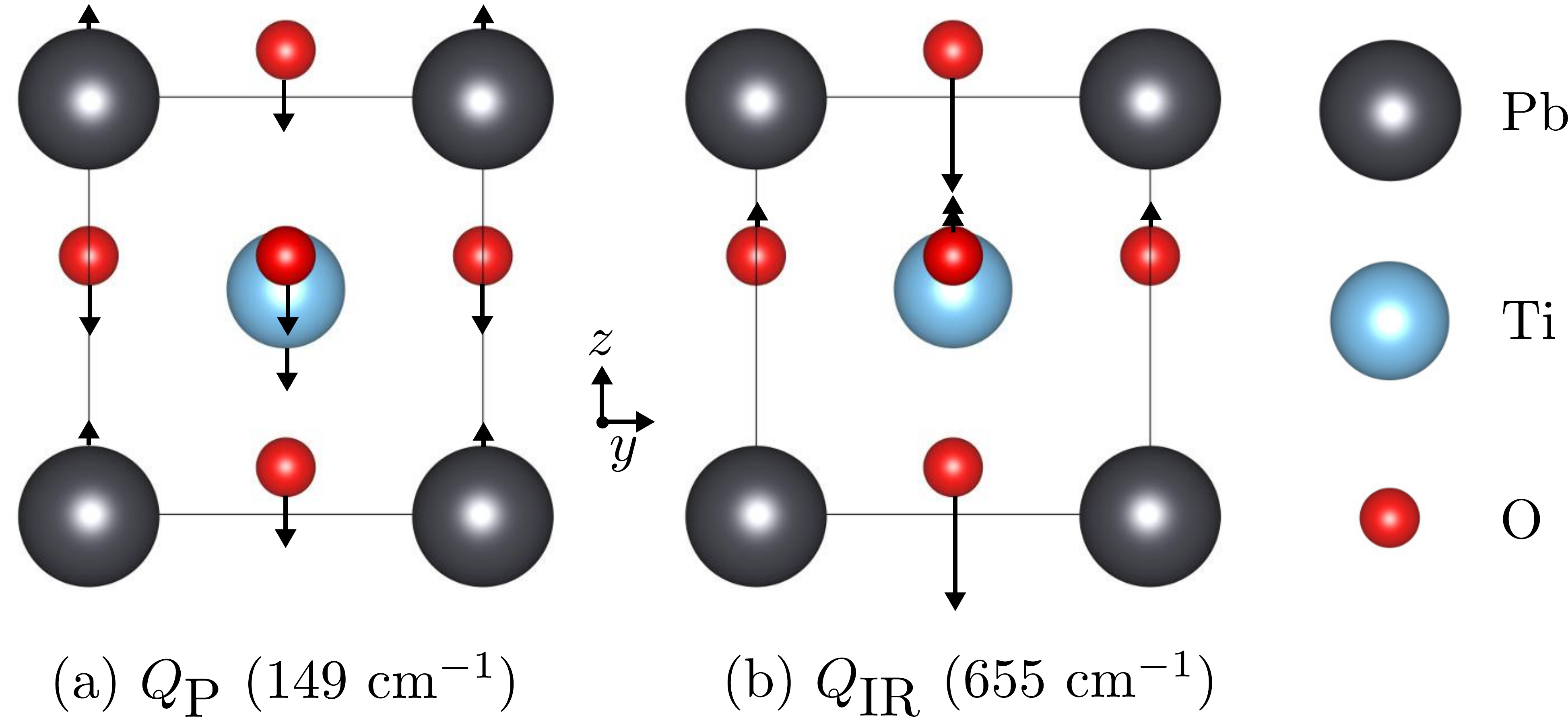}\\
    \includegraphics[width=0.5\columnwidth]{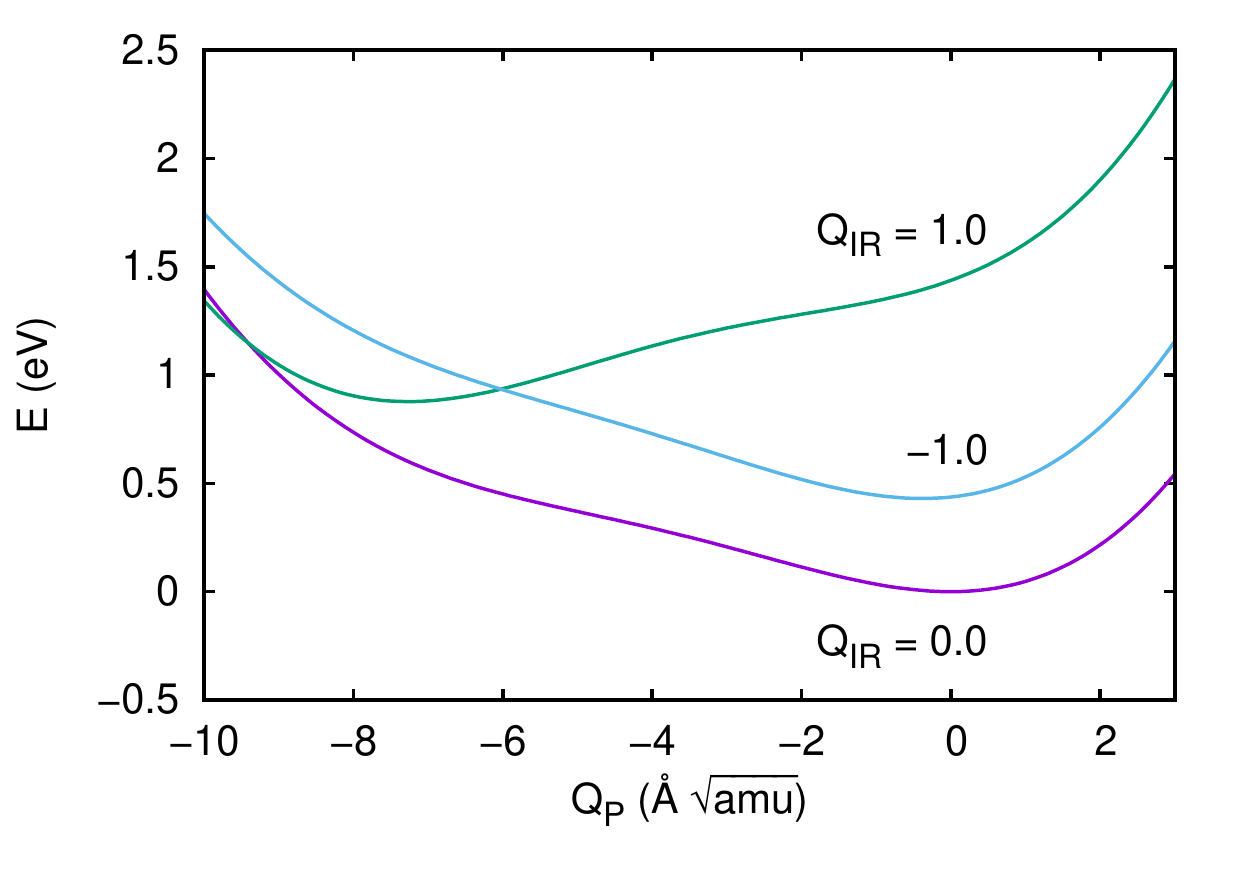}
    \end{center}
  \caption{Top: Displacement patterns of (a) lowest-frequency $\qp$
  and (b) highest-frequency $\qir$ modes of the ferroelectric phase
  of PbTiO$_3$.  Bottom: Total energy as a function of the $\qp$ 
  coordinate for several values of the $\qir$ coordinate. Reproduced from
  \cite{sube15}. }
  \label{fig:pto-phon-vec}
\end{figure}

Although the capability to pump the IR-active phonons of transition metal
oxides has been available since 2007 \cite{rini07}, a mechanism for switching
ferroelectric order using nonlinear phononics was not discussed until
2015 \cite{sube15}.  The main reason for this delay in tackling this problem was a
conceptual misunderstanding.  Since the 1970s, nonlinear phonon couplings
had been discussed in terms of ionic Raman scattering where excited
IR-active phonons couple to Raman phonons via $\qr^{}\qir^2$ or
$\qr \qiro \qirt$ couplings \cite{wall71,mart74}.  In centrosymmetric crystals,
only Raman-active phonons, which do not break inversion symmetry, can
couple to IR-active phonons at this order.  Because the atomic
displacement pattern associated with a ferroelectric order parameter
derives from an unstable IR-active phonon, lattice
displacements that change the ferroelectric order parameter via
nonlinear phononics was not explored.  In 2015, Subedi pointed out
that any phonon mode $\qp$ that modifies the ferroelectric
polarization is both Raman and IR active due to the lack of inversion
symmetry in ferroelectric materials \cite{sube15}.  Thus, a cubic-order
coupling is allowed between $\qp$ and any IR-active phonon that can be
externally pumped.  The question is whether the coupling is large and
causes displacement along the direction that switches the
ferroelectric order. This question was answered in the affirmative for
the case of PbTiO$_3$.

\begin{figure}
  \begin{center}
    \includegraphics[width=0.6\columnwidth]{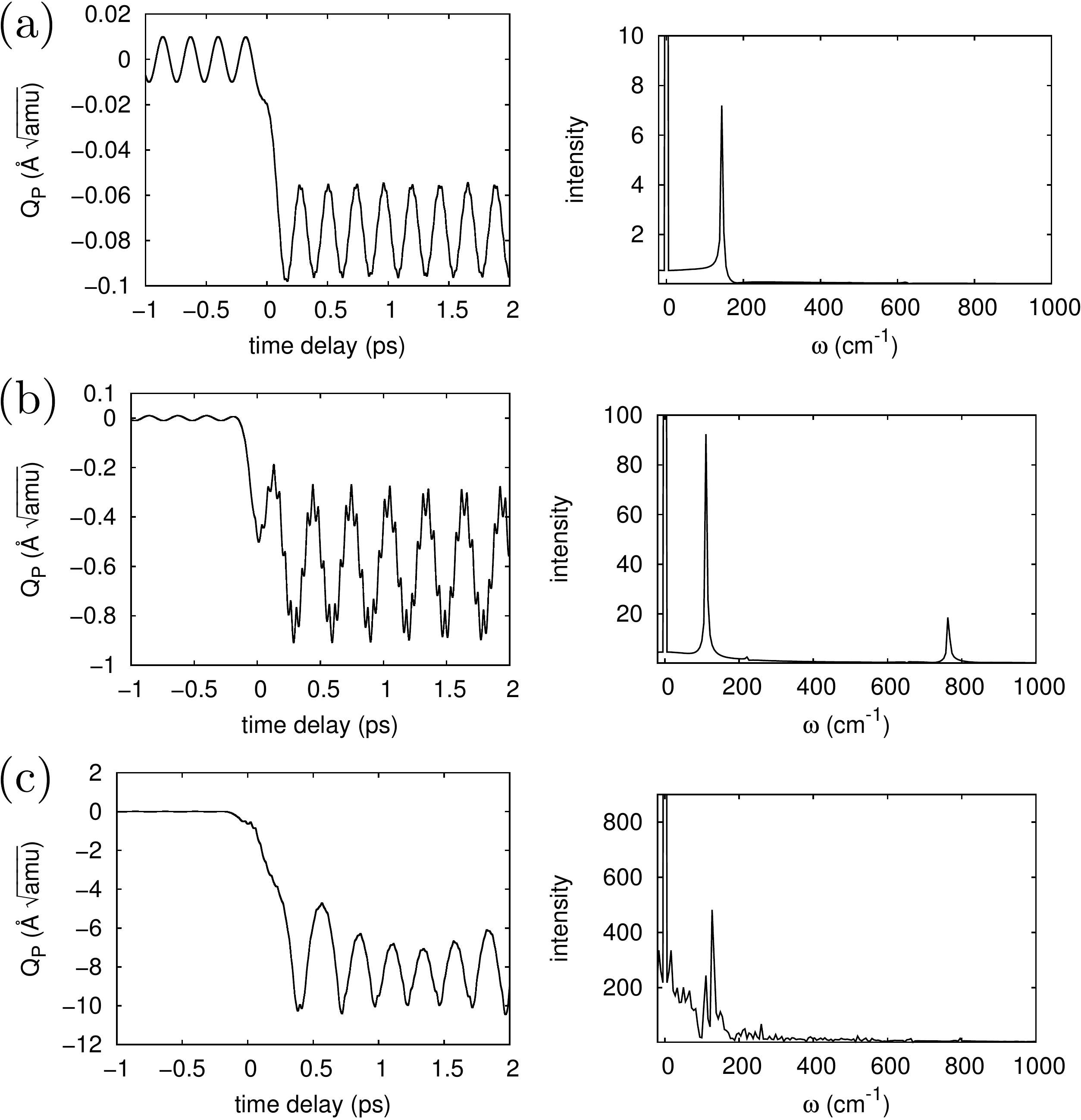}
    \end{center}
  \caption{Dynamics of the $\qp$ mode for three different pump amplitudes. 
    Left panels: Displacements along $\qp$ coordinate as function of time delay. 
    Right panels: Fourier transform of the positive time delay oscillations.  
    Damping effects have been neglected. Reproduced from \cite{sube15}.}
  \label{fig:pto-dyn}
\end{figure}

The displacement patterns of the lowest ($\qp$) and highest ($\qir$)
frequency phonon modes of PbTiO$_3$ are shown in Fig.~\ref{fig:pto-phon-vec}(top).  
The calculated total energy
as a function of the $\qp$ coordinate for several values of the $\qir$
coordinate is shown in Fig.~\ref{fig:pto-phon-vec}(bottom).  The minimum of the
energy curve for the $\qp$ coordinate shifts towards the switching
direction for both positive and negative values of the $\qir$
coordinate, which reflects the presence of a $\qp^{} \qir^2$ nonlinear
coupling term.  There is an asymmetry in energy as a function of the
$\qp$ coordinate because of the presence of a large $\qp^3$
anharmonicity.  Due to this anharmonic term, the minimum of the $\qp$
coordinate suddenly jumps to a large negative when the value of the
$\qir$ coordinate is continuously increased to large positive values.
This causes an abrupt reversal of the ferroelectric polarization without
the magnitude of the polarization getting reduced to a value of zero.
This phenonmena is seen in the numerical solutions of the coupled
equations of motions for the $\qp$ and $\qir$ coordinates in the
presence of an external pump on the $\qir$ mode as shown in Fig.~\ref{fig:pto-dyn}.

% Subedi reported that similar coupling between lowest and highest
% frequency IR-active modes exists in other ferroelectrics such as
% BaTiO$_3$ and LiNbO$_3$.

\begin{figure}
  \begin{center}
    \includegraphics[width=0.5\columnwidth]{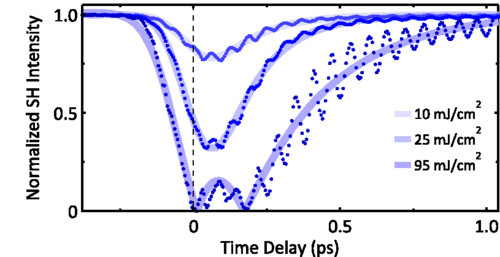}
    \end{center}
  \caption{Time-resolved second-harmonic intensity of LiNbO$_3$ after a mid-IR pump.
    The intensity is normalized to its value before excitation. Reproduced from \cite{mank17}.}
  \label{fig:mank-shg}
\end{figure}

The phenomenon of light-induced switching of ferroelectrics via nonlinear 
phononics proposed by theory was partially confirmed by Mankowsky 
\textit{et al.} \cite{mank17}.   They  performed time-resolved measurements of
SHG intensity and phase of an 800 nm
probe pulse after a mid-IR excitation in LiNbO$_3$ with a pump duration of 150 fs.  
Their result is shown in Fig.~\ref{fig:mank-shg}.  For pump fluences 
smaller than 50 mJ/cm$^2$, the SHG
intensity decreased to a finite value before returning to the
equilibrium value.  Above a threshold fluence of 60 mJ/cm$^2$, the
intensity vanished completely, increased to a finite value, vanished
completely again, and then relaxed to the equilibrium value.  Their
measurement of the phase of the second-harmonic signal showed that the
phase changed by 180$^\circ$ whenever the SHG intensity vanished
completely, which implies a temporary and partial reversal of the
ferroelectric polarization.  Furthermore, as can be seen in Fig.~\ref{fig:mank-shg}, 
the SHG intensity also showed
small modulations corresponding to the oscillations of the pumped IR-active
phonon, and this indicates that the IR-active phonon is coherently
oscillating while the polarization reversal is taking place.  The 
state with switched polarization lasted only for 200 fs. Similar experiments 
with longer pump pulses are necessary to ascertain whether the switching lasts
for the duration of the pump pulse or the pump only causes large-amplitude 
oscillations of the $\qp$ coordinate. In any case, even though
Mankowsky \textit{et al.}\ were not able to permanently switch the electric polarization,
their experiment confirms the theoretical prediction that a cubic-order
nonlinear phonon coupling with a large magnitude and an appropriate sign exists in
oxide ferroelectrics that can reverse the electric polarization.

% \subsection{Effect of Raman-Raman coupling}
% Rondinelli
% Magnetic titanates

\section{Symmetry breaking due to quartic coupling}
\label{}

\subsection{Quartic coupling between a Raman and an IR phonon modes}

Only cubic nonlinearities between phonon modes were discussed in the
literature prior to 2014.  This was presumably because higher order
nonlinearities were thought to be small when the total energy of a
crystal is expanded as a function of phonon coordinates.  Subedi
\textit{et al.}\ pointed out that quartic nonlinearities between two phonon
modes can be large when cubic nonlinearity is not allowed by symmetry \cite{sube14}.
In fact, a $\qr^2\qir^2$ term is the lowest order nonlinearity allowed
by symmetry when $\qr$ has a non-trivial irrep.  

\begin{figure}
  \begin{center}
    \includegraphics[width=0.5\columnwidth]{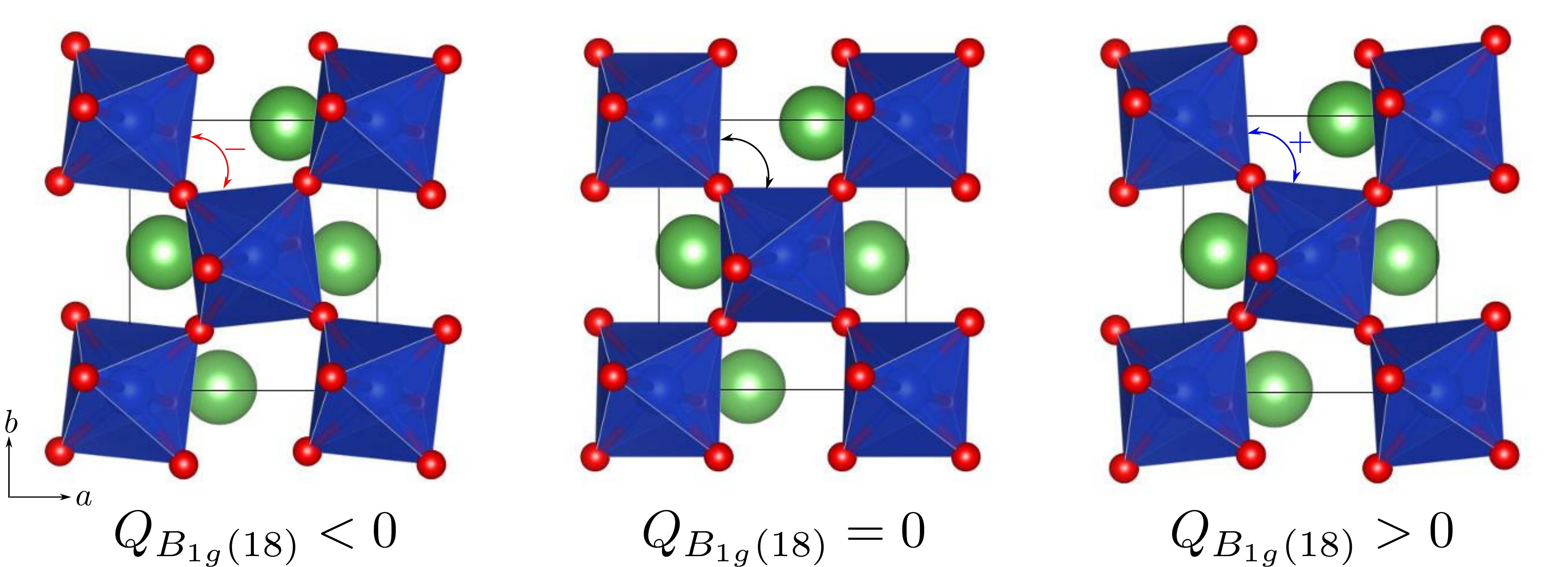}
    \includegraphics[width=0.5\columnwidth]{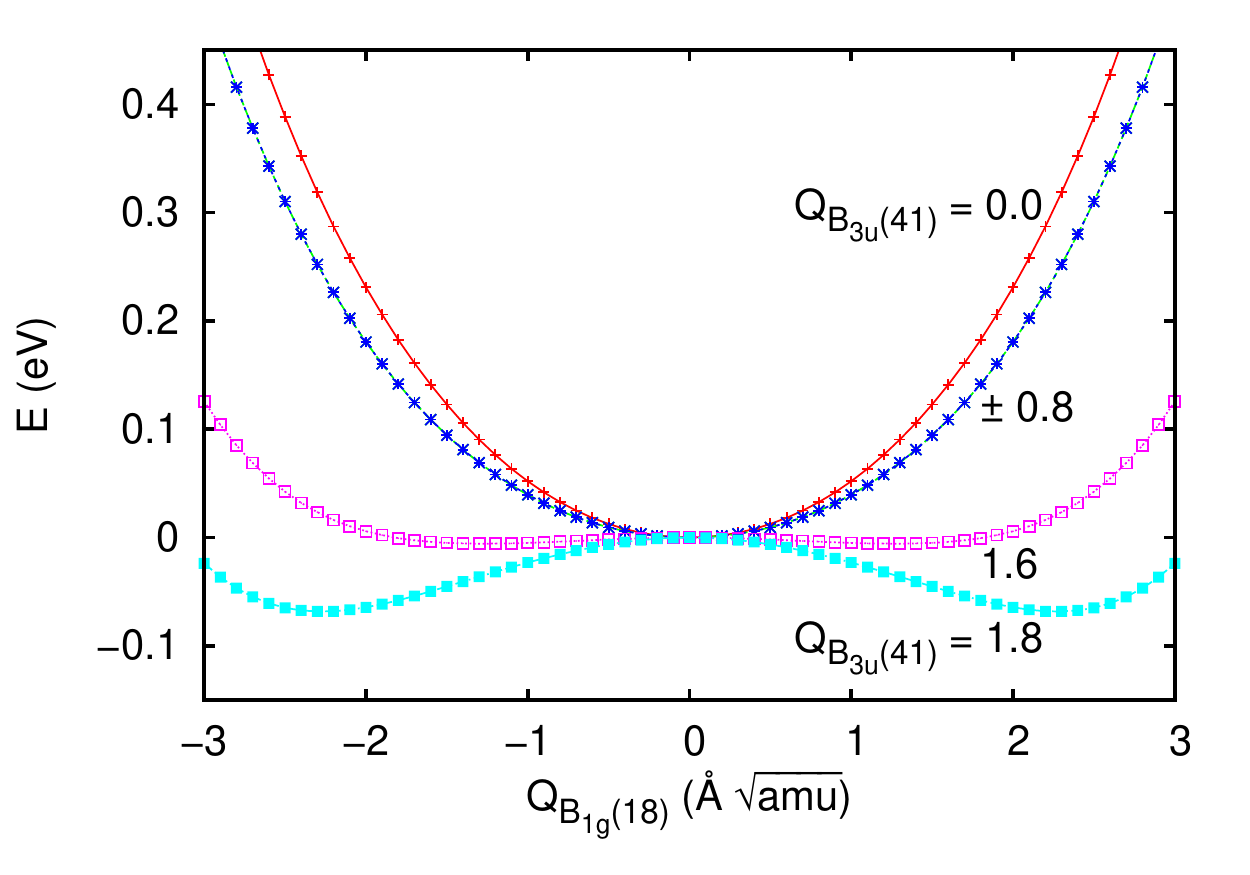}
    \end{center}
    \caption{Top: Sketch of the atomic displacements corresponding to the 
    $\bog(18)$ Raman-active phonon mode of La$_2$CuO$_4$.
    Bottom: Total energy as a function of the $\bog(18)$ 
    coordinate for several values of the $\bru(41)$ coordinate. For visual
    clarity $E(\qr,\qir) - E(0,\qir)$ is plotted so that all curves coincide
    at $\qr = 0$.  Reproduced from \cite{sube14}.}
  \label{fig:lco-ene}
\end{figure}

Such a large quartic order coupling was found in La$_2$CuO$_4$ between its
$\bog$(18) and $\bru$(41) phonon modes \cite{sube14}. The $\bog$(18) mode changes the 
in-plane rotations of the CuO$_6$ octahedra as shown in 
Fig.~\ref{fig:lco-ene}(top), and the $\bru$(41) mode involves
in-plane stretching of the Cu-O bonds.  The $\bog$(18) mode breaks
the $m_x$ and $m_y$ mirror symmetries. Therefore, the structures
generated by the positive and negative values of the $\bog$(18)
coordinates are related by these symmetries, and the total energy is
symmetric as a function of the $\bog$(18) coordinate.  Similarly,
the $\bru$(41) mode breaks the $m_z$ and inversion symmetries, and the
total energy is also symmetric as a function of the $\bru$(41) coordinate.

The calculated total energy surface as a function of the $B_{1g}$(18) and
$B_{3u}$(41) coordinates is shown in Fig.~\ref{fig:lco-ene}(bottom).  It fits the  
expression
\begin{eqnarray}
E(\qr,\qir) & = & \frac{1}{2} \omr^2 \qr^2 + \frac{1}{2} \omir^2\qir^2 
+ \frac{1}{4} a_4 \qr^4 + \frac{1}{4} b_4 \qir^4
\nonumber \\ 
& &-\frac{1}{2} g \qr^2 \qir^2,
\label{eq:energ_lco}
\end{eqnarray}
where $\qr$ and $\qir$ denote the coordinates of the $B_{1g}(18)$ and
$B_{3u}$(41) modes, respectively.  As expected, the calculated energy surface is
even with respect to both $\bog$(18) and $\bru$(41) coordinates,
which is in contrast to the $\qr^{}\qir^2$ nonlinearity that is even only
with respect to the IR phonon coordinate.  The energy curve of the 
$B_{1g}$(18) coordinate softens as the value of the $B_{3u}$(41) coordinate
is increased, and it develops a double well beyond a threshold value of the 
$B_{3u}$(41) coordinate. In Eq.~\ref{eq:energ_lco}, this is reflected by a 
positive value of the coupling coefficient $g$.

\begin{figure}
  \begin{center}
    \includegraphics[width=0.5\columnwidth]{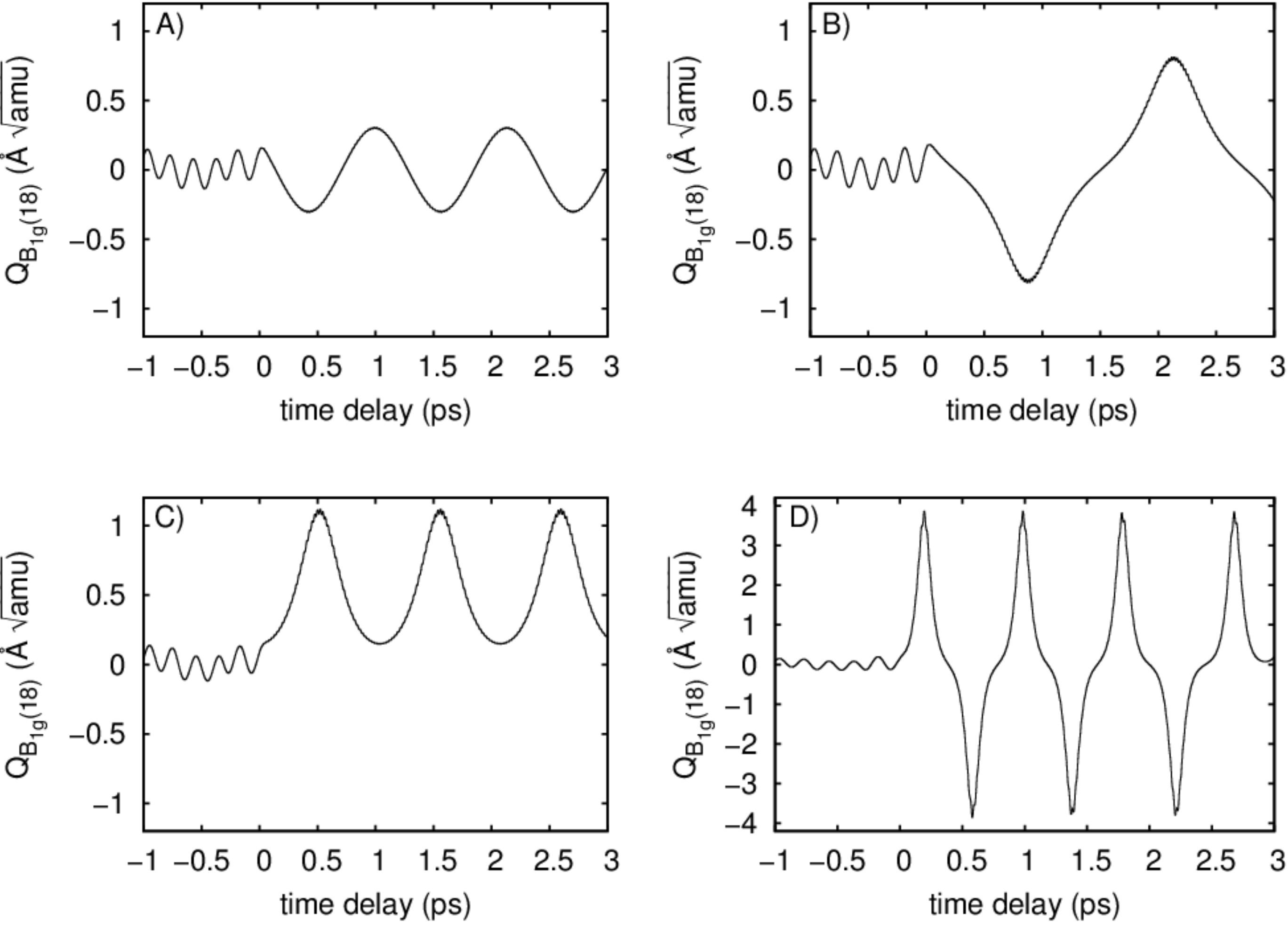}
    \end{center}
  \caption{Dynamics of the $\bog$(18) coordinate due to a quartic nonlinearity in
   La$_2$CuO$_4$.  Damping effects have been neglected.  Reproduced from \cite{sube14}.}
  \label{fig:lco-dyn}
\end{figure}

Since a finite value of the $B_{3u}$(41) coordinate decreases the curvature of 
the energy curve of the $B_{1g}$(18) coordinate, this implies that frequency of 
the $B_{1g}$(18) mode changes while the $B_{3u}$(41) is coherently oscillating.  
Furthermore, 
the $B_{1g}$(18) mode should oscillate at a displaced position at one of the 
local minima of the double-well potential beyond a critical value of the 
amplitue of $B_{3u}$(41) mode.  This
picture was confirmed by solving the coupled equations of motion of these 
modes, which are
\begin{eqnarray}
\ddot{Q}_{\textrm{\small{IR}}}+\omir^2\qir &=& g \qr^2\qir -  b_4 \qir^3 + F(t) \nonumber \\
\ddot{Q}_{\textrm{\small{R}}}+\omr^2\qr&=& \frac{1}{2} g \qr \qir^2 - a_4 \qr^3\,.
\end{eqnarray}
Here $F(t)=F\sin(\Omega t) e^{-t^2/2\sigma^2}$ is the external pump term on
the $B_{3u}$(41) coordinate, and $F$, $\sigma$, and
$\Omega$ are the amplitude, width and frequency of the pump light pulse,
respectively.  Numerical solutions of these equations revealed four qualitatively
different behavior for the oscillations of the $B_{1g}$(18) mode, as depicted in
Fig.~\ref{fig:lco-dyn}. The Raman mode oscillates about its local minimum below a threshold 
value $F_c$ of the external pump [Fig.~\ref{fig:lco-dyn}(a)].  Near this threshold, there is a narrow
range where it makes a long period oscillation about the local maximum of the
double well potential, which is analogous to the Kapitza phenomenon for a vibrating 
pendulum [Fig.~\ref{fig:lco-dyn}(b)].  As the pump amplitude is increased further, it oscillates
at a displaced position in one of the minima of the double well [Fig.~\ref{fig:lco-dyn}(c)].  At even
larger values of the pump amplitude, it again oscillates about the equilibrium
position with a large amplitude that encompasses both minima of the double well
potential [Fig.~\ref{fig:lco-dyn}(d)].  

Since $B_{1g}$(18) mode breaks the $m_x$ and $m_y$ mirror symmetries, the oscillations
about the displaced position in Fig.~\ref{fig:lco-dyn}(c) describe a light-induced dynamical
symmetry breaking of the crystal.  This is a non-perturbative effect that occurs
above a critical threshold of the pump field. Although intense pump pulses with
peak fields greater than 10 MV/cm are available these days, experimental studies
of this phenomenon in La$_2$CuO$_4$ have not yet been reported in the literature.

% symmetry breaking not observed

\subsection{Light induced ferroelectricity due to quartic coupling
  between two IR phonon modes}
  
Quartic nonlinearities of the type $Q_1^2 Q_2^2$ are allowed by symmetry between
any two phonon coordinates $Q_1$ and $Q_2$ because the square of an irrep is the
trivial irrep.  Therefore, two IR-active phonon modes can also couple with
each other. Subedi showed that such a nonlinearity can lead to transiently 
induced ferroelectricity in strained KTaO$_3$ \cite{sube17}.  This phenomenon was illustrated
for the case of 0.6\% compressively strained KTaO$_3$ by studying the dynamics 
of its two lowest-frequency IR-active phonons when its highest-frequency
IR-active phonon mode is pumped.  The two lowest-frequency phonon modes in this
system have the irreps $A_{2u}$ and $E_u$. The $A_{2u}$ mode with calculated 
frequency $\omlz = 20$ cm$^{-1}$ involves 
atomic motions along the $z$ axis, whereas the doubly degenerate $E_u$ mode 
with frequency $\omlx = \omly = 122$ cm$^{-1}$ causes the atoms to move in the 
$xy$ plane.  The highest-frequency mode that should be pumped to induce ferroelectricity has the irrep
$E_u$ with frequency $\omhx = \omhy = 556$ cm$^{-1}$.

% Their calculated frequencies are $\omlz = 20$ and $\omlx
% = \omly = 122$  cm$^{-1}$, respectively.  The highest-frequency mode has the irrep $E_u$ and 
% calculated frequency $\omhx = \omhy = 556$ cm$^{-1}$.

\begin{figure}
  \begin{center}
    \includegraphics[width=0.8\columnwidth]{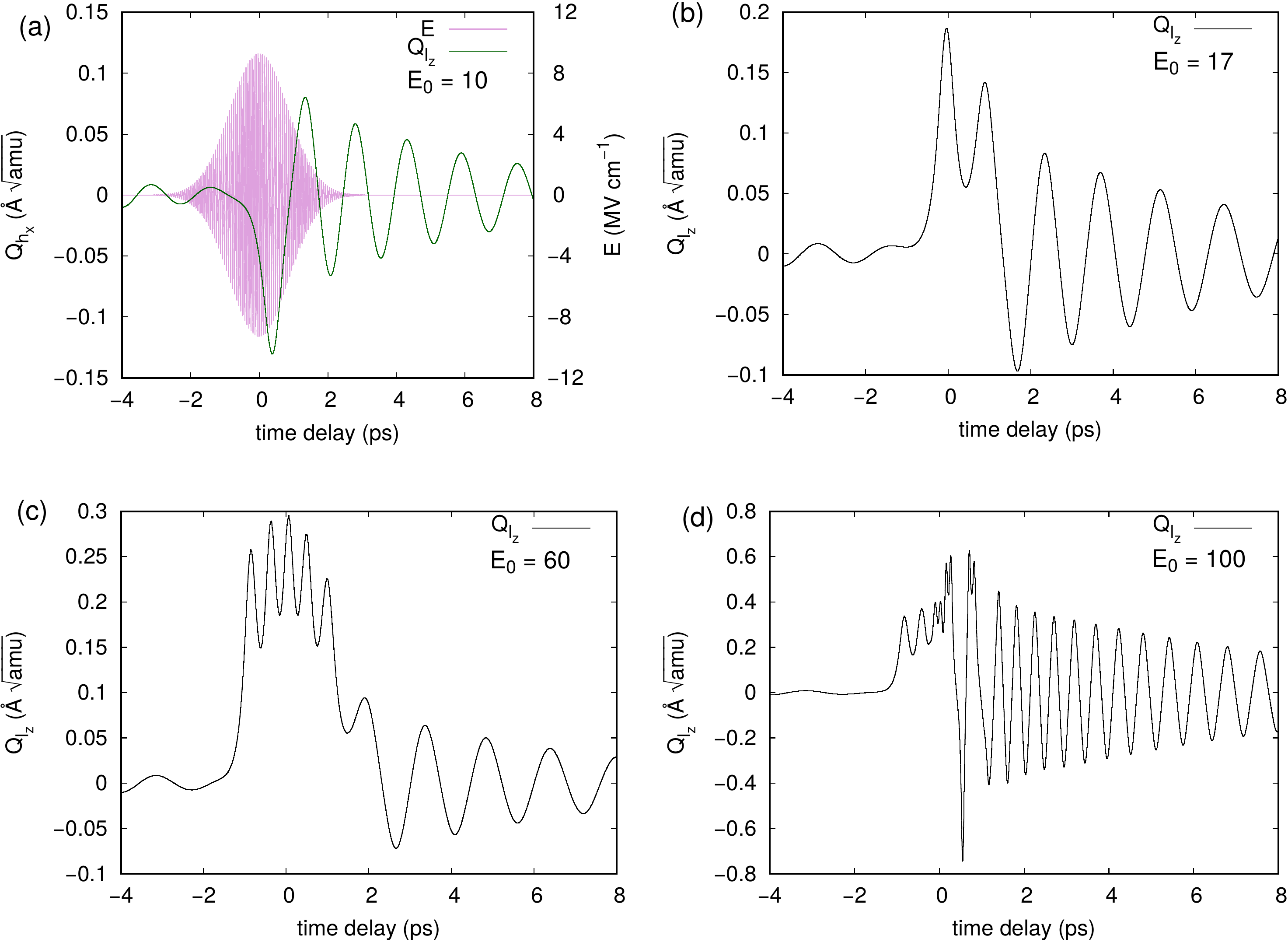}
    \end{center}
  \caption{Dynamics of the $\qlz$ coordinate of strained KTaO$_3$ after 
  $\qhx$ coordinate has been pumped by an external pulse $E$ with duration of 
  2 ps. The dynamics for four different values of the peak electric field $E_0$ 
  (MV cm$^{-1}$) of the pump pulse are shown.  Damping effects are taken into 
  account in this study.  Reproduced from \cite{sube17}.}
  \label{fig:kto-dyn}
\end{figure}

Total energy calculations as a function of the phonon coordinates showed that
 the highest-frequency mode couples to the two lowest-frequency
modes in this system with large quartic nonlinearities. The coupling is such
that the energy curve of the low-frequency $E_u$ coordinate $\qlx$ stiffens, whereas the
energy curve of the low-frequency $A_u$ coordinate $\qlz$ softens when the 
highest-frequency coordinate $\qhx$ has a finite value.  The calculated total energies as a
function of these three coordinates were fit to a polynomial. 
These coordinates were treated as classical oscillators, and the fitted
polynomial was used as their potential energy.  The coupled equations of motion
in the presence of an external pump were numerically solved for several values
of pump intensities, and four such solutions for the $\qlz$ coordinate is shown
in Fig.~\ref{fig:kto-dyn}.  Similar to the case of $\qr^2\qir^2$ coupling in 
La$_2$CuO$_4$ discussed in the previous section, here also the lowest-frequency
$\qp$ coordinate oscillates about a local minimum above a pump threshold [Figs.~\ref{fig:kto-dyn}(b) and (c)].
Since the displacement along the $\qp$ coordinate breaks inversion symmetry,
these calculations show that light-induced ferroelectricity can be stabilized
due to the $\qp^2\qir^2$ nonlinear coupling.  

Nova \textit{et al.}\ have pumped the highest-frequency IR-active phonon of paraelectric
SrTiO$_3$ \cite{nova19}.  They did not observe any second harmonic signal of an optical probe pulse from
the sample after it was pumped by a single mid-IR pulse, indicating that the sample remained paraelectric
after the mid-IR pump.  
However, when the sample was exposed to a mid-IR pump for several minutes, the formation 
of a metastable ferroelectric state was inferred from a finite second harmonic signal of the probe pulse.  
The metastable ferroelectric state persisted for several hours after being exposed to the mid-IR 
irradiation.  Intriguingly, similar metastable ferroelectric state was obtained by using teraherz pump
\cite{li19}.  This suggest that nonlinear phononics may not be the cause of transient 
ferroelectricity in the experiment of Nova \textit{et al}.

\section{Phonon upconversion due to ionic Raman scattering}

Majority of the experimental and theoretical investigations of the nonlinear 
phononics phenomena have focused on pumping the high-frequency IR-active phonons of 
materials to induce dynamics along their low-frequency phonon modes.  
% These phenomena
% require laser sources that are nearly resonant with the frequency of the
% IR-active phonon modes. 
Juraschek and Maehrlein proposed that a phenomenon
analogous to sum frequency Raman scattering can occur due to a $\qiro^{}\qirt^2$
nonlinearity in a material with two IR-active phonons with the relation 
$\omiro =  \omirt / 2 $ \cite{jura18}.  They found that when the low-frequency 
coordinate $\qiro$ is resonantly pumped, the cubic nonlinearity can cause 
oscillations of the high-frequency coordinate $\qirt$.
 
\begin{figure}
  \begin{center}
    \includegraphics[width=0.5\columnwidth]{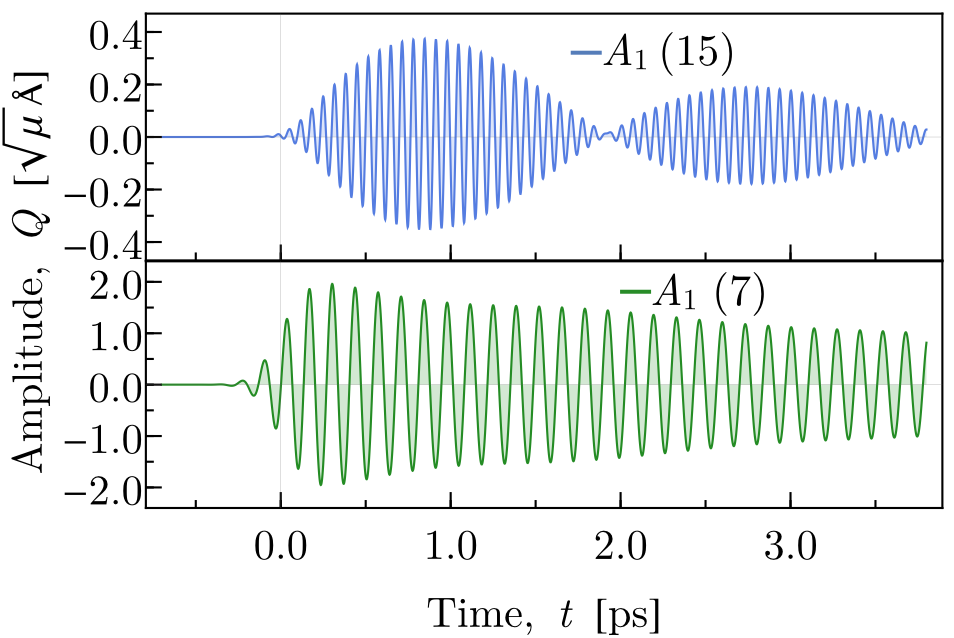}
    \end{center}
  \caption{Dynamics of the high-frequency $\qirt$ coordinate (denoted by $A_1$(15))
  after the low-frequency $\qiro$ coordinate (denoted by $A_1$(7)) 
  is externally pumped. Reproduced from \cite{jura18}.}
  \label{fig:bfo-jura}
\end{figure}

They considered the simplest form of nonlinear lattice potential $V(\qiro,\qirt)
= \frac{1}{2} \omiro^2\qiro^2 + \frac{1}{2} \omirt^2\qirt^2  + 
c \qiro^{}\qirt^2$.  This leads to the coupled equations of motions
\begin{eqnarray}
\ddot{Q}_{\textrm{\small{IR$_1$}}} +
\gamma_{\textrm{\small{IR$_1$}}}\dot{Q}_{\textrm{\small{IR$_1$}}} +
(\omiro^2 + 2c\qirt)\qiro & = & Z_{\textrm{\small{IR$_1$}}} E(t), \nonumber \\
\ddot{Q}_{\textrm{\small{IR$_2$}}} + 
\gamma_{\textrm{\small{IR$_2$}}}\dot{Q}_{\textrm{\small{IR$_2$}}} +
\omirt^2 \qirt & = & c \qiro^2(t).
\end{eqnarray}
Here, $\gamma_{\textrm{\small{IR$_1$}}}$ and $\gamma_{\textrm{\small{IR$_2$}}}$
describe the damping of the $\qiro$ and $\qirt$ coordinates, respectively, and
$Z_{\textrm{\small{IR$_1$}}}$ denotes the mode effective charge of the low-frequency $\qiro$
coordinate. 
%
% In addition, they also consider the excitation of the $\qirt$ mode
% via two-photon absorption process by the equation
% \begin{eqnarray}
% \ddot{Q}_{\textrm{\small{IR$_2$}}} + 
% \gamma_{\textrm{\small{IR$_2$}}}\dot{Q}_{\textrm{\small{IR$_2$}}} +
% \omirt^2 \qirt & = & \varepsilon_0 R E^2(t),
% \end{eqnarray}
% where $\varepsilon_0$ is the vacuum permittivity and $R$ is the Raman tensor. 
%
The results of numerical solutions of these equations in the presence of a
finite driving field $E(t)$ with frequency $\omega_0 = \omiro$ is shown in
Fig.~\ref{fig:bfo-jura}, which shows the high-frequency $\qirt$ mode oscillating due to sum-frequency 
upconversion.

Kozina \textit{et al.}\ have experimentally demonstrated this phenomenon in 
SrTiO$_3$ \cite{kozi19}. When they pumped the lowest-frequency transverse optic TO$_1$ phonon
of this material, they also observed lattice oscillations at higher frequencies 
corresponding to the transverse optic TO$_2$ and TO$_3$ modes in time-resolved x-ray
diffraction experiments.  The TO$_1$ mode has a frequency of 1.5--2.5 THz depending on
the sample temperature, whereas the TO$_2$ and TO$_3$ modes have frequencies of 5.15
and 7.6 THz, respectively.  This indicates that the phonon upconversion mechanism
proposed by Juraschek and Maehrlein works even when the frequencies of the high-frequency phonon 
modes are not integer multiples of the frequency of the pumped low-frequency phonon mode.

\section{Control of magnetism via nonlinear phononics}

\subsection{Magnon excitation via nonlinear magneto-phonon coupling in 
ErFeO$_3$}

ErFeO$_3$ is an antiferromagnetic insulator with a small residual ferromagnetic
moment that arises due to a canting associated with the Dzyaloshinskii-Moriya
interaction.  Earlier in this review, the observation of the $A_g$ and $B_{1g}$ phonon
modes after a mid-IR pump by Nova \textit{et al.}\ was discussed \cite{nova16}. 
Interestingly, they also observed oscillations corresponding to a low-frequency
magnon when the $B_{2u}$ and $B_{3u}$ IR-active phonons were simultaneously excited.
The $B_{2u}$ and $B_{3u}$ modes are polarized along the $b$ and $a$ axes,
respectively. Since a finite value of their respective coordinates $\qwu$ and
$\qru$ leads to a formation of finite electrical dipole moments along the $b$
and $a$ axes, respectively, simultaneous excitation of these modes using a circularly 
polarized pulse should give rise to circulating charges inside the lattice. 
Nova \textit{et al.}\ proposed that this generates an effective magnetic field
that excites the low-frequency magnon.  Their scenario has been supported by a
microscopic theory based on first principles calculations \cite{jura17b}.

\subsection{Modifying the magnetic state of a material}

Nonlinear phononics can coherently modify atomic distances inside a crystal.
In magnetic materials, this can also alter exchange interactions, and a
modified magnetic state might get stabilized in the light-induced transient
state.  Fechner \textit{et al.}\ have theoretically proposed that the equilibrium 
antiferromagnetic ordering of Cr$_2$O$_3$ gets modified to another
antiferromagnetic ordering with ferromagnetically coupled nearest-neighbor
spins when its high-frequency IR-active phonon mode with the irrep $A_u$ is 
externally pumped \cite{fech18}.  This occurs because the Cr-Cr distances increase 
in the transient state as a result of the displacement along an $A_g$ Raman mode due to a
$\qr^{}\qir^2$ nonlinearity. Similar modification of the equilibrium magnetic 
state to a hidden antiferromagnetic state has been proposed in
the rare-earth titantes by Gu and Rondinelli \cite{gu18} and Khalsa and 
Benedek \cite{khal18}. 

More interestingly, Radaelli has proposed that ferroelectricity and
ferromagnetism can be induced in piezoelectric and piezomagnetic materials,
respectively, by simultaneously pumping the orthogonal components $\qir^x$
and $\qir^y$ of a doubly 
degenerate IR-active mode with the irrep $E_u$ \cite{rada18}.  The metastable states occur
because the simultaneous pumping of the orthogonal components causes a displacement
along a Raman-active phonon mode with a nontrival irrep due to a $\qr^{xy}\qir^x \qir^y$
nonlinearity.  It was shown that the displacement along the Raman coordinate $\qr^{xy}$, 
which transforms as $xy$, is given by
\begin{eqnarray}
  \qir^{xy} & \propto & 2 \qirmax^{x} \qirmax^{y} \cos \Delta \phi  \nonumber \\
   & \propto & 2 E_x E_y \cos \Delta \phi.
\end{eqnarray}
Here $\qirmax^x$ and $\qirmax^y$ are the amplitudes of the $\qir^x$ and $\qir^y$
modes, respectively.  $E_x$ and $E_y$ are the magnitudes of the
electric fields used to pump the $\qir^x$ and $\qir^y$ modes, 
respectively, and $\Delta \phi$ is their phase difference. Because the displacement along
the $\qir^{xy}$ coordinate is proportional to $\cos \Delta \phi$, $\qir^{xy}$ has a finite
value only when the orthogonal components of $E_u$ are pumped in-phase or 
out-of-phase.  Furthermore, the displacement along $\qr^{xy}$ coordinate switches direction
wherever the phase difference changes by $\pi$.  Thus, the $\qr^{xy}\qir^x \qir^y$
nonlinearity can induce ferroelectricity or ferromagnetism if the ferroelectric polarization
or ferromagnetic moment is proportional to the $\qr^{xy}$ coordinate, and the direction of the induced
ferroelectric or ferromagnetic moment can be controlled by changing the phase difference of the pump
pulse.  Radaelli has suggested
that peizoelectric BPO$_4$ and piezomagnetic CoF$_2$ are candidate materials where this type of 
ferroelectricity and ferromagnetism can be induced, respectively, using nonlinear phononics.

\begin{figure}
  \begin{center}
    \includegraphics[width=0.8\columnwidth]{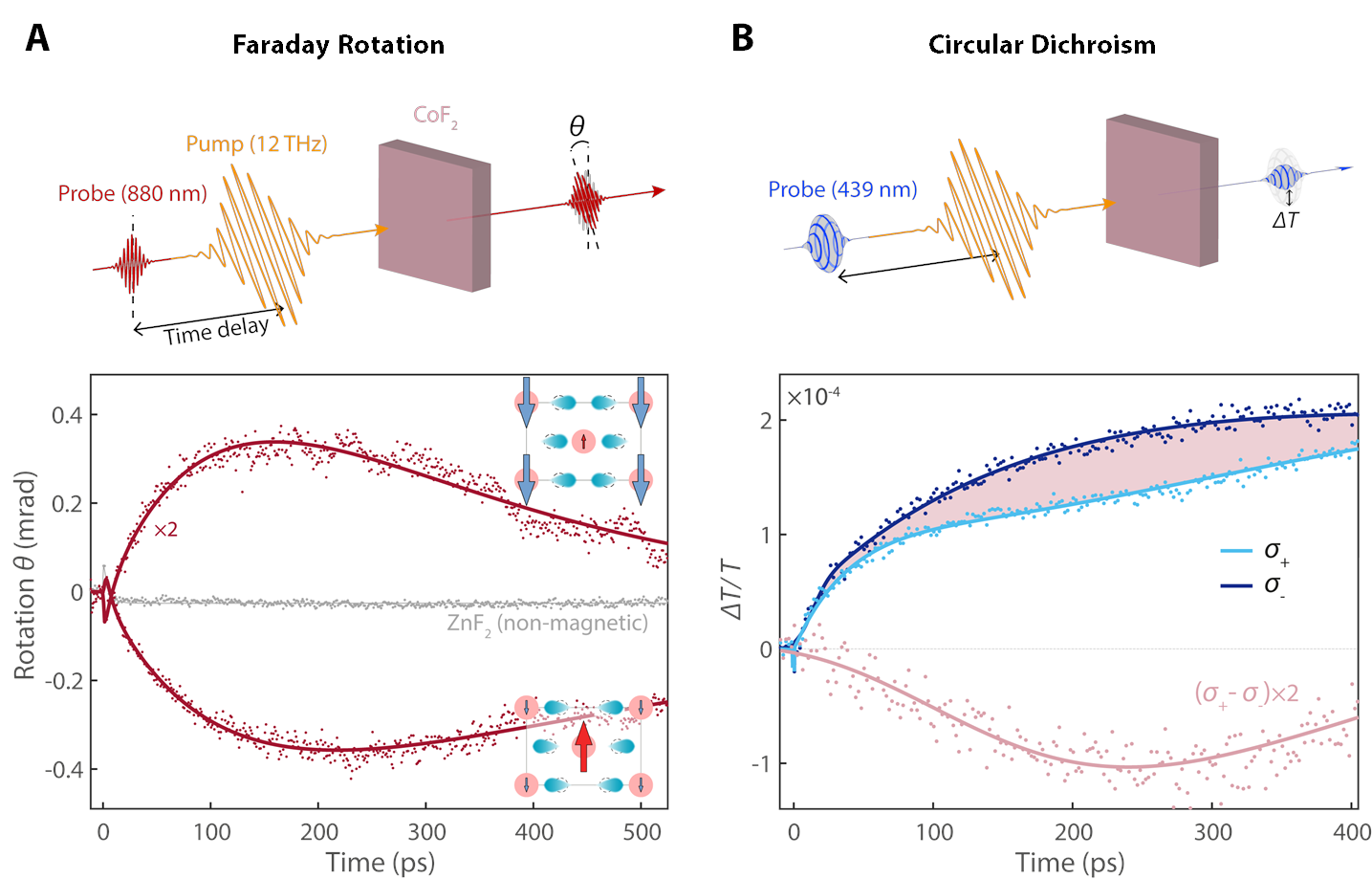}
    \end{center}
  \caption{A(top): Depiction of mid-IR pump-Faraday rotation probe setup. A(bottom):
  Faraday rotation in CoF$_2$ after a 12 THz pump for two polarizations of the pump
  pulses ($+45^{\circ}$ and $-45^{\circ}$. 
  B(top): Depiction of mid-IR pump-circular dichroism probe setup.  B(bottom): The  
  relative change in transmission for left (dark blue) and right (light blue) 
  circular polarized probe pulses. Reproduced from \cite{disa20}. }
  \label{fig:cof2-exp}
\end{figure}

Disa \textit{et al.}\ have recently demonstrated this phenomena in CoF$_2$ 
\cite{disa20}. This
material is a compensated antiferromagnet below $T_N = 39$ K, and an application of
a strain along the [110] direction induces a ferrimagnetic state with a finite
magnetic moment.  They were able to stabilize a similar ferrimagnetic state by 
simultaneously pumping the orthogonal components of its high-frequency IR-active
 phonon with the
irrep $E_u$, which should displace the lattice along a Raman-active phonon with the
irrep $B_{2g}$ due to the $\qr^{xy}\qir^x \qir^y$ nonlinearity.  
A displacement along the $B_{2g}$ mode causes one set of Co-F
distances in the material to lengthen while Co-F distances in another sublattice
shortens, and this is responsible for the uncompensation of Co moments.  The
presence of a finite net magnetic moment in the transient state was confirmed
by time-resolved measurements of the Faraday rotation and circular dicroism
of probe pulses. As shown in Fig.~\ref{fig:cof2-exp}, a pump-induced magnetic
signal was immediately observed, which changed sign after 7 ps.  After the sign reversal,
the magnetic signal continued to grow until it reached its maximum value at
200 ps.  The reason for such a long-lived magnetic signal has not been
completely understood.  One possibility proposed by Disa \textit{et al.}\ is 
that the transient displacement along the Raman-active phonon coordinate is
reinforced by induced magnetic moment.  Time-resolved x-ray diffraction studies
should help in understanding this long-lived metastable state by clarifying
the nature of the structural distortions in the light-induced phase.  In particular,
oscillations and displacements along the $B_{2g}$ coordinate should be observed while
the $\qir^x$ and $\qir^y$ coordinates are simultaneously oscillating to confirm that
the transient ferrimagnetism is due to nonlinear phononics.

\section{Conclusions}
% \label{}

In summary, nonlinear phononics is an emerging field that has the potential 
to develop as a powerful method for controlling materials by stabilizing novel crystal
structures that cannot be accessed in equilibrium.  This is made possible by
coherent atomic displacements along a set of phonon coordinates after a selective
excitation of the IR-active phonons of a material, and it contrasts with the incoherent
atomic motions that result from heating.  Nonlinear coupling of the pumped IR-active 
phonon to other phonons is the microscopic mechanism 
responsible for the coherent lattice displacement.  Intense mid-IR pump pulses are
now available, and mid-IR light-induced control of materials properties have been
demonstrated in pump-probe experiments.  Nevertheless, this field is still in infancy
compared to the pump-probe experimental activities that are performed in chemistry
laboratories. 2D pump-probe spectroscopy experiments are routinely used by chemists
to directly observe simultaneous excitations of the pumped vibrational mode as well as 
other modes that are excited due to nonlinear couplings.  The nonlinearity between
different vibrational modes are reflected by the presence of off-diagonal signals in
2D spectroscopy, and they can be used to quantify the nonlinear couplings. Mid-IR 
pump-second harmonic probe experiments similar to 2D spectroscopy have been recently
performed on the wide band gap insulator LiNbO$_3$, and they have demonstrated simultaneous 
oscillations of the pumped IR-active mode while the lattice gets displaced along a 
Raman-active phonon coordinate. Several mid-IR pump induced phase transitions have 
been attributed to coherent lattice displacements due to nonlinear phononics, including
insulator-metal transitions and melting of spin and orbital orders.  Mid-IR pump-induced
increase in reflectivity have also been reported in several superconductors, and they have
been interpreted as signatures of light-enhanced superconductivity.  However,  
excitations of the pumped mode have not been experimentally demonstrated in these 
experiments.  More experimental studies that directly measure the nonlinear phonon 
couplings between the pumped phonon and other active phonon degrees of freedom would put 
this field on a stronger footing.

A microscopic theory based on first principles calculations of nonlinear phonon couplings
has been developed to study the dynamics of a material when its IR-active phonons are
selectively pumped.  In addition to the cubic nonlinearities discussed in the 1970s, 
quartic nonlinearities with large coupling coefficients that can stabilize a symmetry-broken 
phase beyond a threshold value of the pump intensity has been found using this theoretical 
approach.  Theoretical studies have also proposed light-induced reversal of ferroelectric
polarization, ferroelectricity in paraelectrics and ferromagnetism in antiferromagnets, and 
these predictions have been partially confirmed by experiments. Cavity control of nonlinear  
couplings and phono-magneto analog of opto-magneto effect have been shown to be feasible by
calculations.  Their experimental realizations would confirm that nonlinear phononics is truly
a novel way to control the physical properties of materials.

% The Acknowledgements are also a un-numbered section
\section*{Acknowledgements}
% Acknowledgements text here

I am grateful to Antoine Georges and Andrea Cavalleri for previous collaborations
on this subject.  I have also benefited from helpful discussions with Michael F\"orst, 
Roman Mankowsky, Tobia Nova, Matteo Mitrano, Srivats Rajasekaran and Yannis Laplace
on this topic.

% The next command determines the bibliography style. Please do not
% change this.
\bibliographystyle{crunsrt}

%This calls all references from the .bib
\nocite{*}

%  This inserts the bib file
% \bibliography{samplebib}

\end{document}